\documentclass[12pt]{article}
\usepackage{amssymb,amsmath}
\usepackage{epsfig}
\usepackage{epstopdf}
\usepackage{latexsym}
\def\ds{\displaystyle}

\begin{document}

\begin{titlepage}

\begin{flushright}
NSF-KITP-09-49
\end{flushright}

\bigskip
\bigskip
\bigskip
\bigskip
\bigskip
\centerline{\Large \bf Deformations of Holographic Duals}
\medskip
 \centerline{\Large \bf to
Non-Relativistic CFTs}
\bigskip
\bigskip
\centerline{{\bf Nikolay Bobev$^{1,2}$ and Arnab Kundu$^2$}}

\bigskip
\centerline{${}^{1}$ Kavli Institute for Theoretical Physics}
\centerline{University of California Santa Barbara}\centerline{
Santa Barbara CA 93106-4030, USA}
\bigskip
\centerline{$^2$ Department of Physics and Astronomy}
\centerline{University of Southern California} \centerline{Los
Angeles, CA 90089, USA}
\bigskip
\centerline{{\rm bobev@usc.edu, akundu@usc.edu} }
\bigskip \bigskip

\begin{abstract}

We construct the non-relativistic counterparts of some well-known
supergravity solutions dual to relevant and marginal deformations
of $\mathcal{N}=4$ super Yang-Mills. The main tool we use is the
null Melvin twist and we apply it to the $\mathcal{N}=1$ and
$\mathcal{N}=2^*$ Pilch-Warner RG flow solutions as well as the
Lunin-Maldacena solution dual to $\beta$-deformations of
$\mathcal{N}=4$ super Yang-Mills. We also obtain a family of
supergravity solutions with Schr\"{o}dinger symmetry interpolating
between the non-relativistic version of the $\mathcal{N}=1$
Pilch-Warner and Klebanov-Witten fixed points. A generic feature
of these non-relativistic backgrounds is the presence of
non-vanishing internal fluxes. We also find the most general,
three-parameter, null Melvin twist of the ${\rm AdS}_5\times S^5$
black hole. We briefly comment on the field theories dual to these
supergravity solutions.

\end{abstract}

\topmargin=-0.8in \oddsidemargin=-0.4in \textheight=9.0in
\textwidth=6.8in

\end{titlepage}

\topmargin=-0.4in \oddsidemargin=-0.2in \textheight=8.8in
\textwidth=6.8in







\section{Introduction and Motivation}

The gauge/gravity duality is a powerful tool for understanding
strongly coupled field theories and it has been applied
extensively to a plethora of relativistic conformal field theories
in various dimensions \cite{Aharony:1999ti,D'Hoker:2002aw}. In
condensed matter physics there exist strongly coupled systems
which respect the non-relativistic analog of the conformal group,
known as the Schr\"{o}dinger group. Recently it has been shown
that one can realize certain gravity solutions which are invariant
under the Schr\"{o}dinger group and thus extend the gauge/gravity
duality to non-relativistic systems. This provides a new
opportunity to apply holographic techniques to condensed matter
systems\footnote{See \cite{Son:2008ye}-\cite{Kachru:2008yh} for a
sample of recent work and \cite{Hartnoll:2009sz} for a review and
a more complete list of references.}.

The celebrated example of gauge/gravity duality (in its weak form)
is the equivalence between classical type IIB supergravity in
AdS$_5\times S^5$ background and the large 't Hooft coupling limit
of the planar $\mathcal{N}=4$ super Yang-Mills (SYM) theory. It
has been further realized that this correspondence can be extended
to a large class of deformations of $\mathcal{N}=4$ SYM which
possess less global symmetry and supersymmetry. These include
relevant deformations \cite{Freedman:1999gp, Freedman:1999gk}
triggered by turning on certain mass terms for the adjoint
scalars\footnote{See \cite{D'Hoker:2002aw} for a review on
holographic RG flows and a more complete list of references.} and
the marginal $\beta$-deformations \cite{Leigh:1995ep,Lunin:2005jy}
corresponding to deformation of the $\mathcal{N}=4$
superpotential. It seems natural to look for non-relativistic
cousins of these gravity solutions and this will be our goal in
this note.

Gravity solutions invariant under the Schr\"{o}dinger group were
first constructed in five dimensions
\cite{Son:2008ye,Balasubramanian:2008dm} and later embedded in ten
dimensions by performing a null Melvin twist on the familiar ${\rm
AdS}_5\times S^5$ solution
\cite{Herzog:2008wg,Maldacena:2008wh,Adams:2008wt}. The null
Melvin twist \cite{Gimon:2003xk,Hubeny:2005qu} is a solution
generating technique which can be applied to any solution of
ten-dimensional supergravity with one compact and one non-compact
$U(1)$ isometries. It consists of performing a boost, T-duality,
shift, T-duality and another boost. If the compact manifold has at
least $U(1)^3$ isometry (which is the case of $S^5$) one can
generalize this transformation by allowing three independent
shifts along the three independent $U(1)$ directions. Using this
we obtain the most general null Melvin twist of the ${\rm
AdS}_5\times S^5$ black hole solution. When all three shifts are
set equal the background reduces to the one found in
\cite{Herzog:2008wg,Maldacena:2008wh,Adams:2008wt} and at zero
temperature the background is invariant under the full
Schr\"{o}dinger group.

We also exploit the null Melvin twist to generate holographic
duals to the non-relativistic version of $\mathcal{N}=4$ SYM
deformed by relevant and marginal operators. There are two RG
flows that have explicit ten-dimensional gravity duals. The
solution of \cite{Pilch:2000fu} corresponds to an $\mathcal{N}=1$
supersymmetric RG flow triggered by a mass term for one of the
three complex chiral superfields of $\mathcal{N}=4$ SYM. This
solution flows to an $\mathcal{N}=1$ superconformal fixed point in
the IR and the gravity dual of this conformal fixed point was
originally found in \cite{Pilch:2000ej}. The other type IIB
background that we consider is dual to the $\mathcal{N}=2^*$
supersymmetric RG flow induced by turning on masses for two of the
chiral superfields of $\mathcal{N}=4$ SYM. The gravity dual of
this flow was found in \cite{Pilch:2000ue} (see also
\cite{Brandhuber:2000ct}). In the infrared it does not flow to a
fixed point, instead there is a particular distribution of D3
branes which is interpreted as the Coulomb branch of the gauge
theory\footnote{We will make a slight abuse of language here and
use the term non-relativistic Coulomb branch flow to describe the
non-relativistic version of the Coulomb branch RG flow of the SYM.
}. We also find the non-relativistic generalizations of the one
parameter family of gravity solutions interpolating between the
Klebanov-Witten(KW) point \cite{Klebanov:1998hh} and a
$\mathbb{Z}_2$ orbifold of the Pilch-Warner(PW)
\cite{Pilch:2000ej} fixed point \cite{Halmagyi:2005pn}. Finally we
consider the non-relativistic cousin of the Lunin-Maldacena
solution \cite{Lunin:2005jy} dual to the $\beta$-deformation of
$\mathcal{N}=4$ SYM. All these solutions have at least one $U(1)$
isometry on the compact manifold and we find their
non-relativistic versions by applying the null Melvin twist along
this $U(1)$. An important general feature of the solutions of
\cite{Lunin:2005jy,Pilch:2000fu,Pilch:2000ue,Halmagyi:2005pn} is
the presence of internal NS and RR flux on the internal manifold.
The null Melvin twist preserves these fluxes and to the best of
our knowledge the solutions in Section 4 and Section 7 are the
first examples of gravity backgrounds with Schr\"{o}dinger
isometry which have non-trivial internal flux.

It is worth mentioning that an RG flow geometry which flows from a
relativistic fixed point to a non-relativistic Lifshitz-like fixed
point has been constructed in \cite{Kachru:2008yh}. In contrast,
the RG flow background discussed in Section 3 interpolates between
two vacua invariant under the Schr\"{o}dinger symmetry.

It should be noted that there is no known finite temperature
version of the ten-dimensional $\mathcal{N}=1$ and
$\mathcal{N}=2^{*}$ PW solutions, so we refrain from addressing
finite temperature aspects of the geometries dual to the
non-relativistic RG-flows\footnote{There was some work on the
finite temperature $\mathcal{N^{*}}=2$ PW solution in
\cite{Buchel:2003ah}, however the explicit ten dimensional
solution is not known.}. The finite temperature Lunin-Maldacena
solution is well known and we find its non-relativistic version by
applying the null Melvin twist. It will be interesting to analyze
the thermodynamics of this background and understand how the
marginal deformation affects certain transport coefficients in the
dual field theory.

We start in Section 2 with a review of the Schr\"{o}dinger
symmetry and its dual realization in type IIB supergravity, we
also construct the most general null Melvin twist of ${\rm
AdS}_5\times S^5$. We introduce the PW background in Section 3 and
in section 4 show that the Melvin twisted PW background possesses
the Schr\"{o}dinger symmetry in the UV and the IR. Section 5
contains the non-relativistic version of the family of fixed
points interpolating between the $\mathbb{Z}_2$ orbifold of the PW
fixed point and the KW fixed point. In Section 6 we discuss the
non-relativistic version of the $\mathcal{N}=2^{*}$ Coulomb branch
RG flow solution. We apply the null Melvin twist to the
Lunin-Maldacena background in Section 7 and in Section 8 we
present some comments about the field theories dual to the
geometries that we construct via the null Melvin twist. Section 9
contains some concluding remarks and open problems. Various
technical details are presented in the Appendices.

\section{Schr\"{o}dinger symmetry}

\subsection{The algebra}

The symmetry group of the Schr\"{o}dinger equation in flat space
is known as the Schr\"{o}dinger symmetry. The corresponding
algebra is generated by spatial translations $P^i$, temporal
translation $H$, spatial rotations $M^{ij}$, Galilean boost $K^i$,
the dilatation operator $D$, a special conformal transformation
$C$ and the Galilean mass $M$. The explicit form of the algebra is
given by the following commutation relations ($i=1,...,d$, where
$d$ is the number of spatial dimensions)
\begin{eqnarray}
&&[M^{ij},M^{kl}] = i \left( \delta^{ik}M^{jl} +\delta^{jl}M^{ik}
- \delta^{il}M^{jk} - \delta^{jk}M^{il}\right)~, ~~~
[M^{ij},P^{k}] = i \left(\delta^{ik}P^{j}-
\delta^{jk}P^{i}\right)~, \notag\\
&&[M^{ij},K^{k}] = i \left(\delta^{ik}K^{j}-
\delta^{jk}K^{i}\right)~, ~~~ [D,P^{i}]=-iP^{i}~, ~~~
[D,K^{i}]=iK^{i}~, ~~~ [D,H]=-2iH\,\\
&& [P^{i},K^{j}] = -i\delta^{ij}M~, ~~~ [D,C]=2iC~, ~~~
[H,C]=iD~.\notag
\end{eqnarray}
The last two commutation relations involving the special conformal
transformation generator $C$ can only be included when the
dynamical exponent, characterizing the different scaling of space
and time, is 2. In this case, also the commutator $[D,M]=0$.
Therefore the states are simultaneous eigenstates of the
dilatation and the mass operator.

The Schr\"{o}dinger algebra in $d$-spatial dimensions can be
obtained by the light cone reduction of the relativistic conformal
algebra in $(d+2)$-spatial dimensions. This can be intuitively
understood by noticing that the light cone reduction of the
massless Klein-Gordon equation (which is conformal) gives the
Schr\"{o}dinger equation in free space, we refer to
\cite{Son:2008ye,Balasubramanian:2008dm} for further details. Here
we will be interested in the case $d=2$, i.e. non-relativistic
field theories in $2+1$ dimensions.

\subsection{The dual geometry}

In \cite{Son:2008ye, Balasubramanian:2008dm}, a corresponding
five-dimensional gravitational background was constructed which
possesses the Schr\"{o}dinger isometry group in two spatial
dimensions. It was further realized in \cite{Herzog:2008wg,
Maldacena:2008wh, Adams:2008wt} that it is possible to embed this
geometry in ten dimensions. This can be done by applying a
solution generating technique, the null Melvin twist, to the
well-known ${\rm AdS}_5\times S^5$ background (in the Poincare
patch) \cite{Gimon:2003xk,Hubeny:2005qu}. Here we will apply the
most general null Melvin twist and generalize the background found
in \cite{Herzog:2008wg, Maldacena:2008wh, Adams:2008wt}.

We start with the planar ${\rm AdS}_5\times S^5$ non-extremal
black hole solution (The radii of the ${\rm AdS}_5$ and $S^5$ are
equal to $L$)
\begin{equation}
ds^2 = L^2 r^2 [ -F(r) dt^2 +dy^2 + dx_1^2+dx_2^2 ] +
\ds\frac{L^2}{r^2F(r)} dr^2+
L^2\ds\sum_{i=1}^{3}(d\mu_i^2+\mu_i^2d\varphi_i^2)~,
\label{adsBHmet}
\end{equation}
\begin{equation}
F_{(5)} = L^4 (  r^3 dt\wedge dx_1\wedge dx_2\wedge dy \wedge dr +
\sin^3\vartheta\cos\vartheta\sin\xi\cos\xi d\vartheta\wedge d\xi
\wedge d\varphi_1 \wedge d\varphi_2 \wedge
d\varphi_3)~,\label{adsBHflux}
\end{equation}
where
\begin{equation}
\mu_1 = \cos\vartheta~, \qquad\qquad \mu_2 =
\sin\vartheta\cos\xi~, \qquad\qquad \mu_3=\sin\vartheta\sin\xi~,
\nonumber
\end{equation}
and
\begin{equation}
F(r) = 1 - \ds\frac{r_{+}^4}{r^4}~.\nonumber
\end{equation}
The coordinates on $S^5$ are chosen such that the $U(1)$
isometries along $\phi_i$ are the $U(1)^3$ Cartan subgroup of
$SO(6)$. Now we can apply the null Melvin twist to this
background. The procedure is straightforward to implement and
amounts to the following operations: first we boost in the $(t,y)$
plane with parameter $\gamma_0$, then we perform a T-duality along
$y$, then we shift all three Cartan angles of $S^5$ by $\varphi_i
\to \varphi_i + a_i y$, then we perform another T-duality along
$y$ and finally we perform an inverse boost in the $(t,y)$ plane
with parameter $-\gamma_0$ and take the limit $a_i\to 0$,
$\gamma_0\to \infty$ such that $a_i
\cosh\gamma_0=a_i\sinh\gamma_0=\text{finite}$. We also define
$\eta_i\equiv a_i\cosh\gamma_0=a_i\sinh\gamma_0$. Note that we are
doing something a bit more general than the transformation in
\cite{Herzog:2008wg, Maldacena:2008wh, Adams:2008wt} where the
case $a_1=a_2=a_3$ was considered which corresponds to null Melvin
twist along the Hopf fiber of $S^5$. One can think of the third
step of the null Melvin twist as three simultaneous shifts in all
three Cartan directions (or a TsssT transformation for short
\cite{Lunin:2005jy,Frolov:2005dj}). This general null Melvin twist
generates a metric of the form
\begin{multline}
ds^2 = L^2r^2 \left[-
\ds\frac{r^2F(r)q(\vartheta,\xi)}{K(r)}(dt+dy)^2 -
\ds\frac{F(r)}{K(r)}dt^2 + \ds\frac{dy^2}{K(r)}
+dx_1^2+dx_2^2\right]+\ds\frac{L^2}{r^2F(r)} dr^2 \\+
L^2\ds\sum_{i=1}^{3}\left[ d\mu_{i}^2 + \mu_i^2d\varphi_i^2\right]
- \ds\frac{L^2r_{+}^4}{r^2K(r)}
\left(\ds\sum_{i=1}^{3}L^2\eta_i\mu_i^2d\varphi_i\right)^2~,
\label{general NMT met}
\end{multline}
as well as an NS two-form and a non-trivial dilaton given by
\begin{equation}
B_{(2)} = \ds\frac{L^2r^2}{K(r)}
\left(\ds\sum_{i=1}^{3}\eta_i\mu_i^2d\varphi_i\right) \wedge
(F(r)dt+dy)~, \qquad\qquad \Phi(r) = - \ds\frac{1}{2}\log K(r)~,
\label{general NMT fields}
\end{equation}
where we have defined\footnote{For generic values of $\eta_i$ the
function $K$ depends on $\vartheta$ and $\xi$, for brevity we will
denote it just by $K(r)$.}
\begin{equation}
K(r) = 1 + \ds\frac{r_{+}^4}{r^2} q(\vartheta,\xi)~, \qquad
\text{and} \qquad q(\vartheta,\xi) =L^4
\ds\sum_{i=1}^{3}\eta_i^2\mu_i^2~.\nonumber
\end{equation}
The self-dual five-form flux $F_{(5)}$ remains unaffected. This is
the most general null Melvin twist of the non-extremal D3 brane
solution of which we can now take various limits.

First let us consider the zero temperature (extremal) limit which
amounts to setting $r_{+}=0$. The background simplifies to
\begin{eqnarray}
&& ds^2=-\frac{L^2q(\vartheta,\xi)}{z^4} du^2+\frac{L^2}{z^2}\left(-2dudv+dx_1^2+dx_2^2+dz^2\right)+L^2\ds\sum_{i=1}^{3}(d\mu_i^2+\mu_i^2d\varphi_i^2)\ ,\nonumber\\
&& B_{(2)}= \frac{L^2}{z^2} \left(\sum_{i=1}^3  \eta_i \mu_i^2
d\varphi_i\right)\wedge du ~ , \label{genNMTextreme}
\end{eqnarray}
where we have defined new coordinates
\begin{equation}
u=t+y\ ,\qquad v=\frac{1}{2}\left(t-y\right)~, \qquad z =
\ds\frac{1}{r}~.
\end{equation}
The dynamical exponent of the solution, $\nu$, parametrizes the
different scaling of time and space in the dual non-relativistic
theory and is determined by the $g_{uu}$ term in the metric,
$g_{uu}\sim z^{-2\nu}$. Our solutions have $\nu=2$, this is
expected since only non-relativistic systems with $\nu=2$ admit
the full Schr\"{o}dinger symmetry. Indeed, following
\cite{Son:2008ye}, one can verify that the background above has
the full Schr\"{o}dinger symmetry. The generators of the
Schr\"{o}dinger algebra are given by the following isometries of
the metric ($\epsilon$ and $\epsilon_i$ ($i=1,2$) are
infinitesimal parameters)
\begin{eqnarray}
P^i &:& x_i \to x_i + \epsilon_i~, \qquad H: u \to u + \epsilon~,
\qquad M: v \to v + \epsilon \ ,\notag\\\notag\\
K^i &:& x_i \to x_i - \epsilon_i u~, \qquad v \to v -
\epsilon_i x_i~, \qquad M_{12}: x_1 \to x_1 + \epsilon x_2~, \qquad x_2 \to x_2 - \epsilon x_1\ , \notag\\\label{Schgen}\\
D &:& x_i \to (1-\epsilon)x_i ~, \qquad z \to (1-\epsilon)z~,
\qquad u \to (1-\epsilon)^2 u ~, \qquad v \to v\ ,\notag\\\notag\\
C &:& x_i \to (1-\epsilon u)x_i ~, \qquad z \to (1-\epsilon u)z~,
\qquad  u \to (1-\epsilon u) u ~, \qquad v \to v -
\ds\frac{\epsilon}{2}(x_ix_i + z^2 )\ .\notag
\end{eqnarray}
One can easily check that the zero temperature background above is
invariant under these infinitesimal transformations. Alternatively
one can show that the five-dimensional non-compact metric has nine
Killing vectors and their Lie brackets close under the
Schr\"{odinger} algebra. It should be possible to construct
generalizations of the solution (\ref{genNMTextreme}) with
different dynamical exponents, $\nu>2$, along the lines of
\cite{Hartnoll:2008rs}, the field theories dual to such solutions
will not be symmetric under special conformal transformations
since these are present only for $\nu=2$.

It is interesting to consider also the case
$\eta=\eta_1=\eta_2=\eta_3$, then the extremal solution simplifies
even further to
\begin{eqnarray}
&& ds^2=-\frac{L^6\eta^2}{z^4}
du^2+\frac{L^2}{z^2}\left(-2dudv+dx_1^2+dx_2^2+dz^2\right)\\&&~~~~~~~~~~~~~+L^2\left(d\mu^2
+\ds\frac{\sin^2\mu}{4} (\sigma_1^2 + \sigma_2^2) +
\ds\frac{\sin^2\mu\cos^2\mu}{4} \sigma_3^2 +
\left(d\psi + \ds\frac{\sin^2\mu}{2}\sigma_3\right)^2\right)\ ,\nonumber\\
&& B_{(2)}= \frac{L^2\eta}{z^2} \left(d\psi +
\ds\frac{\sin^2\mu}{2}\sigma_3\right)\wedge du ~ ,
\end{eqnarray}
where the metric on $S^5$ has been written as a Hopf
fiber\footnote{We refer to Appendix D for the explicit coordinate
change.} over $\mathbb{CP}^2$, $\sigma_i$ are the left invariant
$SU(2)$ one forms
\begin{eqnarray}\label{eqt: sigma}
\sigma_1 &=& \cos\alpha_3 ~ d\alpha_1 + \sin\alpha_1 ~\sin\alpha_3
~d\alpha_2~,\notag\\
\sigma_2 &=& \sin\alpha_3 ~ d\alpha_1 - \sin\alpha_1 ~\cos\alpha_3
~d\alpha_2~,\label{sigmas}\\
\sigma_3 &=& d\alpha_3 + \cos\alpha_1 ~d\alpha_2~,\notag
\end{eqnarray}
and
\begin{equation}
J = \ds\frac{1}{2}d\mathcal{A} \equiv \ds\frac{1}{4}~
d\left(\sin^2\mu~\sigma_3\right)\nonumber
\end{equation}
is the K\"{a}hler form on $\mathbb{CP}^2$. This is the type IIB
background constructed in
\cite{Herzog:2008wg,Maldacena:2008wh,Adams:2008wt} and as we
already emphasized it is a special case of the more general null
Melvin twist of ${\rm AdS}_5\times S^5$ given in (\ref{general NMT
met}), (\ref{general NMT fields}).

The Galilean mass in the Schr\"{o}dinger algebra, $M$, can be
thought of as the number density in the non-relativistic field
theory and is identified with the momentum along the compact
$v$~direction,~$P_{v} = \frac{M}{R_{v}}$, where $R_{v}$ is the
radius of the $v$ circle
\cite{Herzog:2008wg,Maldacena:2008wh,Adams:2008wt}. The black hole
solution after the twist has entropy, temperature and a chemical
potential conjugate to $P_v$. In addition to that we have the
three deformation parameters $\eta_i$ representing the freedom to
choose the twist parameters for the three possible $U(1)$
isometries which are global symmetries in the dual field theory.
The overall scale associated with these three twist parameters is
related to the chemical potential. In the case in which all three
$\eta_i$ vanish we have a DLCQ of ${\rm AdS}_5$ (or ${\rm
AdS}_5$-BH) corresponding to zero number density - i.e., no
particles in the non-relativistic theory~\cite{Adams:2008wt}.

An important feature of the general null Melvin twist is the
appearance of the function~$q(\vartheta,\xi)$ in front of $du^2$.
For generic values of $\eta_i$ this function is positive definite
and thus there are no singularities or causal pathologies
\cite{Marolf:2002bx, Brecher:2002bw}. However for special choices
of $\eta_i$, $q(\vartheta,\xi)$ may have zeroes. We have checked
explicitly that the curvature of the ten dimensional solution is
finite for all values of $\eta_i$ which implies that the zeroes of
$q(\vartheta,\xi)$ are not physical singularities and the ten
dimensional solution is regular\footnote{We thank Mukund Rangamani
for useful comments on this point.}. Functions which depend on the
internal manifold do appear in the $g_{uu}$ component of the
metric in some of the solutions analyzed in
\cite{Hartnoll:2008rs}, however in \cite{Hartnoll:2008rs} these
are eigenfunctions of the Laplacian on the internal manifold and
thus necessarily change sign, which may lead to problems with
stability and causality.

The zero temperature background in equations (\ref{adsBHmet}),
(\ref{adsBHflux}) before the general null Melvin twist has 32
Killing spinors (and thus 32 supercharges) and one can show that
for generic values of $\eta_i$ none of these Killing spinors is
preserved by the non-relativistic background given in equations
(\ref{general NMT met}), (\ref{general NMT fields}). However there
are special values of $\eta_i$  for which some supercharges are
preserved~\cite{progress}.

\section{The Pilch-Warner flow}

We start with the Pilch-Warner flow solution presented in
\cite{Pilch:2000fu,Corrado:2001nv}. We use the metric in
\cite{Corrado:2001nv} since it is written in a more convenient
way. Note that for convenience we make a shift of $\sigma_3$ as
compared to \cite{Corrado:2001nv}, namely $\sigma_{3}^{\rm here}
\to \sigma_{3}^{\rm there}+ d\phi$~. This makes the coordinate
$\phi$ the $U(1)$ R-symmetry direction and the vielbein is
\begin{eqnarray}
e^{\mu+1} &=& \Omega e^{A} dx^{\mu}~, \qquad \mu = 0,1,2,3~,\notag\\
e^{5} &=& \Omega dr~,\notag\\
e^{6} &=& L \ds\frac{\Omega\rho^2}{X_1} \left[ \left(1 -
\ds\frac{3}{2} \cos^2\theta\right) d\phi + \ds\frac{1}{2}
\cos^2\theta \sigma^3 \right] ~,\notag\\
e^{7} &=& L \ds\frac{\Omega}{\rho\cosh\chi} d\theta ~,\\
e^{8} &=& L \ds\frac{\Omega}{\rho\cosh\chi} \sin\theta\cos\theta
\left[\left( \ds\frac{3}{2} + \ds\frac{1-\rho^6}{X_1}\left(1 -
\ds\frac{3}{2} \cos^2\theta\right)\right) d\phi -
\ds\frac{1}{2}\left(1 -
\ds\frac{1-\rho^6}{X_1} \cos^2\theta \right)\sigma_3 \right] ~,\notag\\
e^{9} &=& L \ds\frac{\rho}{2\Omega} \cos\theta~ \sigma_1 ~,\notag\\
e^{10} &=& L \ds\frac{\rho}{2\Omega} \cos\theta ~\sigma_2 ~,\notag
\end{eqnarray}
where
\begin{equation}
X_1 = \cos^2\theta + \rho^6 \sin^2\theta ~,\nonumber
\end{equation}
\begin{equation}
\Omega = \ds\frac{(\cosh\chi)^{1/2}X_1^{1/4}}{\rho^{1/2}} ~.
\end{equation}
The functions $\rho(r)$ and $\chi(r)$ are the two supergravity
scalars that trigger the flow and $\sigma_{i}$ are the $SU(2)$
left-invariant one forms explicitly written in (\ref{eqt: sigma}).
There is a non-zero complex two-form potential
\begin{equation}
\mathcal{B} = \ds\frac{i}{2} \sinh\chi (e^{7} - ie^{8})\wedge
(e^{9} - i e^{10})~,\nonumber
\end{equation}
where the NS and RR 2-forms are given by
\begin{equation}
B_{(2)} = {\rm Re}(\mathcal{B})~, \qquad\qquad C_{(2)} = {\rm
Im}(\mathcal{B})~.
\end{equation}
The self-dual five-form flux is given by
\begin{equation}
F_{(5)} = (1+\star) dx^0\wedge dx^1\wedge dx^2\wedge dx^3 \wedge
(w_r(r,\theta)dr + w_{\theta}(r,\theta) d\theta)~,
\end{equation}
where
\begin{equation}
w_r(r,\theta) = \ds\frac{e^{4A}}{4L}~
\ds\frac{\cosh^2\chi}{\rho^4}~ [ (\cosh(2\chi) - 3)\cos^2\theta +
\rho ^6 (2\rho^6 \sinh^2\chi\sin^2\theta + \cos(2\theta) -3) ]
~,\nonumber
\end{equation}
\begin{equation}
w_{\theta}(r,\theta) =   \ds\frac{e^{4A}}{8\rho^2} ~[2 \cosh^2\chi
+ \rho^6(\cosh(2\chi) - 3) ]\sin(2\theta) ~.
\end{equation}
The scalar flow has two critical points:

\bigskip

1) ${\rm AdS}_5\times {\rm S}^5$, this is the UV critical point
given by $\chi = 0$, $\rho = 1$ and $A = \ds\frac{r}{L} \equiv
\ds\frac{r}{L_{UV}}$.

\bigskip

2) ${\rm AdS}_5\times X_{PW}$, this is the IR critical point
(Pilch-Warner fixed point) and is given by $\chi =
\ds\frac{\log(3)}{2}$, $\rho = 2^{1/6}$ and $A =
\ds\frac{2^{5/3}}{3} \ds\frac{r}{L} \equiv \ds\frac{r}{L_{IR}} $.

\bigskip

Note that as required by the holographic c-theorem
\cite{Freedman:1999gp} the radius of ${\rm AdS}_5$ decreases along
the flow
\begin{equation}
\ds\frac{L_{IR}}{L_{UV}} = \ds\frac{3}{2^{5/3}} \approx
0.944941~.\nonumber
\end{equation}
The metric of this RG flow solution is $SU(2)\times U(1)_{\phi}
\times U(1)_{\alpha_3}$ invariant, however the complex two form
breaks this down to $SU(2)\times U(1)_{\phi}$ because
\begin{equation}\label{eqt: pwb}
\mathcal{B} \sim (\sigma_1 - i \sigma_2) \sim e^{-i \alpha_3}~.
\end{equation}
Since the background along the entire flow is manifestly
$SU(2)\times U(1)_{\phi}$ invariant one can easily apply the null
Melvin twist along the two U(1) isometries $y\equiv x_3$ and
$\phi$. This is done explicitly in Appendix A.

It is worth noting that $U(1)_{\phi}$ is not the Hopf fiber and
one can show that at the UV fixed point the particular null Melvin
twist that we apply to the PW solution corresponds to the
following choice of the twist parameters
\begin{eqnarray}
\eta_1=\eta\ , \qquad \eta_2=-\frac{\eta}{2}=\eta_3\ .
\end{eqnarray}
The full RG flow solution after the null Melvin twist is not
particularly illuminating so we present its explicit form in
Appendix A and we proceed with a discussion of the
non-relativistic version of the flow geometry at the fixed points.

\section{Fixed  points and the Schr\"{o}dinger symmetry}

The supergravity scalar flow after the twist still has two fixed
points and as we show in this section the supergravity backgrounds
at these fixed points posses the full Schr\"{o}dinger symmetry.

\subsection{UV fixed point}

As mentioned in the previous section, at the UV fixed point we
have
\begin{equation}
\rho =1~, \qquad \chi = 0~, \qquad \Omega =1~, \qquad X_1=1~,
\qquad A = \ds\frac{r}{L}~.
\end{equation}
The metric, after the null Melvin twist, takes the form
\begin{multline}
ds^2_{10} = - L^2 e^{4r/L} \left(1 -
\ds\frac{3}{4}\cos^2\theta\right)[\eta (dt+dy)]^2-
e^{2r/L}(dt-dy)(dt+dy)\\ +  e^{2r/L} (dx_1^2 +dx_2^2) + dr^2 + L^2
ds^2_{UV}~,\nonumber
\end{multline}
where $ds^2_{UV}$ is the metric on the $S^5$ and is given by
\begin{equation}
ds^2_{UV} = d\theta^2 +\ds\frac{\cos^2\theta}{4} (\sigma_1^2 +
\sigma_2^2) + \ds\frac{\sin^2\theta\cos^2\theta}{4}\left( 3d\phi -
\sigma_3\right)^2 +
\left(\left(1-\ds\frac{3}{2}\cos^2\theta\right) d\phi +
\ds\frac{\cos^2\theta}{2}\sigma_3\right)^2 ~.\label{PWUVmet}
\end{equation}
The NS two-form is
\begin{equation}
B_{(2)} = L^2 e^{2r/L} \left(\left(1 -
\ds\frac{3}{4}\cos^2\theta\right)d\phi -
\ds\frac{\cos^2\theta}{4}~\sigma_3 \right)\wedge [\eta(dt+dy)]~.
\end{equation}
Now define new coordinates
\begin{equation}
 u = (t+y)~, \qquad v = \ds\frac{1}{2 L^2}(t-y)~, \qquad z = e^{-r/L}~,\qquad \hat{x}_1 =L^{-1}x_1~,\qquad \hat{x}_2 =L^{-1}x_2 ~.\nonumber
\end{equation}
In this coordinate the metric and the B-field becomes
\begin{equation}\label{eqt: pwmelvin}
ds^2_{10} =  L^2\left( - \left(1 -
\ds\frac{3}{4}\cos^2\theta\right)\ds\frac{\eta^2}{z^4}
du^2~+~\ds\frac{1}{z^2}\left( -2dudv ~+~ d\hat{x}_1^2
+d\hat{x}_2^2 ~+~ dz^2\right)\right) ~+~ L^2ds^2_{UV}~.\nonumber
\end{equation}
\begin{equation}
B_{(2)} = \ds\frac{\eta L^2}{z^2} \left(\left(1 -
\ds\frac{3}{4}\cos^2\theta\right)d\phi -
\ds\frac{\cos^2\theta}{4}~\sigma_3 \right)\wedge du~.
\end{equation}
The RR sector in the UV is completely invariant under the null
Melvin twist.

Note the appearance of a function of $\theta$ in front of the
$du^2$ term in the metric. As noted before, this can be traced
back to the fact that we did not use the Hopf fiber of $S^5$ for
the null Melvin twist. It is obvious that the function in front of
the $du^2$ term in the metric above is negative definite.
Therefore there are no spacetime pathologies related to causality
and instability of this background \cite{Marolf:2002bx,
Brecher:2002bw}. As expected the dynamical exponent of this
solution is 2.

\subsection{IR fixed point}

Now we move to the analysis of the IR fixed point. This fixed
point is given by
\begin{equation}
\rho =2^{1/6}~, \qquad \chi = \ds\frac{\log3}{2}~, \qquad \Omega
=\ds\frac{2^{5/12}}{3^{1/4}}(1+\sin^2\theta)^{1/4}~, \qquad
X_1=1+\sin^2\theta~, \qquad A = \ds\frac{r}{L_{IR}}~,\nonumber
\end{equation}
where
\begin{equation}
L_{IR} = \ds\frac{3}{2^{5/3}}L~.
\end{equation}
The metric has the form
\begin{multline}\label{eqt: pwir}
ds^2_{10} = \Omega^2L^2_{IR}\left[-
\ds6\left(\ds\frac{1+3\sin^4\theta}{1+\sin^2\theta}\right)
\ds\frac{\eta^2}{z^4} du^2~+~\ds\frac{1}{z^2}\left(-2dudv ~+~
d\hat{x}_1^2 +d\hat{x}_2^2 ~+~ dz^2\right)\right.
\\ \left.- 2\eta\ds\frac{\cos^2\theta\sin\theta}{1+\sin^2\theta}~\sigma_1\ds\frac{du}{z^2}\right] +ds^2_{PW}~,
\end{multline}
where again we have defined new coordinates
\begin{equation}
u = \frac{2^{4/3}}{3^2} (t+y)~, \quad v =
\frac{3^2}{2^{4/3}}\ds\frac{1}{2 L_{IR}^2}(t-y)~, \quad z =
e^{-r/L_{IR}}~,\quad \hat{x}_1 =L_{IR}^{-1}x_1~,\quad \hat{x}_2
=L_{IR}^{-1}x_2 ~,\nonumber
\end{equation}
and $ds^2_{PW}$ is the metric on the deformed $S^5$ at the
Pilch-Warner fixed point\footnote{This fixed point solution was
originally found (in slightly different coordinates) in
\cite{Pilch:2000ej}, see also \cite{Cvetic:2000tb}.} and is given
by
\begin{multline}
ds^2_{PW} = L^2_{IR}
\ds\frac{2^{\frac{5}{6}}}{3^{\frac{3}{2}}}(1+\sin^2\theta)^{\frac{1}{2}}
\left[ 2 d\theta^2 +
\ds\frac{\cos^2\theta}{1+\sin^2\theta}(\sigma_1^2+\sigma_2^2) +
8\ds\frac{\sin^2\theta\cos^2\theta}{(1+\sin^2\theta)^2}\left(d\phi
- \ds\frac{\sigma_3}{2}\right)^2 \right.
\\+\left.\ds\frac{16}{3}\ds\frac{1}{(1+\sin^2\theta)^2} \left( \left(1 -
\ds\frac{3}{2}\cos^2\theta\right)d\phi+
\ds\frac{\cos^2\theta}{2}\sigma_3\right)^2 \right] ~.
\end{multline}
We again note that the function in front of the $du^2$ term in the
metric is negative definite, hence the IR fixed point is also free
of spacetime pathologies. The NS two-form is
\begin{equation}
B_{(2)} = B_1 + B_{PW}~,\nonumber
\end{equation}
where $B_1$ is the piece of the B-field generated by the null
Melvin twist
\begin{equation}\label{eqt: bnoncom}
B_{1} = \eta L^2_{IR} \ds\frac{2^{\frac{7}{3}}}{3}
\left(\ds\frac{1+3\sin^4\theta}{1+\sin^2\theta} ~d\phi -
\ds\frac{\cos^2\theta}{1+\sin^2\theta}~\sigma_3 \right) \wedge
\ds\frac{du}{z^2}~,
\end{equation}
and $B_{PW}$ is the usual internal B-field of the Pilch-Warner
fixed point solution \cite{Pilch:2000ej}
\begin{equation}
B_{PW} = \ds\frac{2^{\frac{4}{3}}}{3^2}L^2_{IR}\left[
\ds\frac{\cos^2\theta\sin\theta}{1+\sin^2\theta}\left(d\phi -
\ds\frac{1}{2}\sigma_3\right)\wedge \sigma_1 +
\ds\frac{\cos\theta}{2}d\theta \wedge \sigma_2\right]~.
\end{equation}
The five-form RR flux is modified by the null Melvin twist and
takes the form
\begin{equation}\label{eqt: fnoncom}
\widetilde{F}_{(5)}=F_{(5)}+ \eta L^2_{IR}
\ds\frac{2^{\frac{7}{3}}}{3}(1+\star)\frac{du}{z^2}\wedge\ds
\left(\ds\frac{1+3\sin^4\theta}{1+\sin^2\theta} ~d\phi -
\ds\frac{\cos^2\theta}{1+\sin^2\theta}~\sigma_3 \right)\wedge
dC_{(2)}~ .
\end{equation}
The RR two-form $C_{(2)}$ remains unchanged.

It can be checked that background metric (\ref{eqt: pwir}), the NS
two-form (\ref{eqt: bnoncom}) and the RR five-form (\ref{eqt:
fnoncom}) at the IR fixed point are invariant under the
Schr\"{o}dinger symmetry \eqref{Schgen}. However there is a new
interesting feature. Recall that the null coordinate $u$ is
identified with the time coordinate for the non-relativistic field
theory. Therefore the off-diagonal element in the metric between
$u$ and $\sigma_1$ can be interpreted as a rotation along the
compact $\sigma_1$-direction. The presence of this term can be
traced back to the fact that we had non-zero NS flux in the
solution before the twist. It will be interesting to understand
the meaning of this rotation term from the point of view of the
dual field theory. It will be quite interesting to see if this
Schr\"{o}dinger invariant ten-dimensional solution can be reduced
to five dimensions and understood as a solution to some effective
five-dimensional equations of motion. It is not immediately clear
to us that this is possible and we will think of the whole RG flow
solution and the IR fixed point as purely ten-dimensional.

The metric of the relativistic Pilch-Warner flow geometry has an
$U(1)$ isometry which rotates $\sigma_1$ in to $\sigma_2$ (this is
the $U(1)_{\alpha_3}$), however, this is not a symmetry of the
background because of the non-zero two-form (\ref{eqt: pwb}). The
null Melvin twist enhances the breaking of this $U(1)_{\alpha_3}$
by generating an off-diagonal term in the metric. This enhanced
symmetry breaking is present everywhere along the flow and at the
IR fixed point but vanishes in the UV.

The dynamical exponent of this solution is $\nu=2$, so we see that
along the RG flow the dynamical exponent is invariant. It should
be possible to find similar gravity solutions with different
dynamical exponents along the lines of \cite{Hartnoll:2008rs}. To
do this one should make a more general ansatz with $g_{uu}\sim
z^{-2\nu}$ and solve the equations of motion. Due to the presence
of internal fluxes this will be a non-trivial task.

\section{Family of non-relativistic fixed points}

There is a one parameter family of supergravity fixed point
solutions, \cite{Halmagyi:2005pn}, which interpolate between the
$\mathbb{Z}_2$ orbifold of the PW fixed point \cite{Pilch:2000ej}
and the ${\rm AdS}_5\times T^{(1,1)}$ solution discussed by
Klebanov-Witten in \cite{Klebanov:1998hh} and originally found by
Romans \cite{Romans:1984an}. These solutions are gravity duals to
a family of $\mathcal{N}=1$ conformal quiver gauge theories which
can be thought of as IR fixed points of mass deformations of a
$\mathbb{Z}_2$ orbifold of $\mathcal{N}=4$ SYM. An interesting
feature of the interpolating family is that the axion-dilaton
vanishes at both the PW and the KW fixed points but has a
non-trivial dependence on the coordinates of the internal manifold
for all interpolating solutions. Here we will apply the null
Melvin twist to the family of solutions in \cite{Halmagyi:2005pn}
and generate a new family of non-relativistic fixed points
invariant under the Schr\"{o}dinger symmetry. We should note that
the analytic form of this family of solutions is not known,
however in \cite{Halmagyi:2005pn} numerical solutions to the
supersymmetry equations were found and it was shown explicitly
that they interpolate between the KW and PW solutions. More
details on the solutions of \cite{Halmagyi:2005pn} are given in
Appendix B\footnote{Note that we use``hats",e.g. $\hat{f}_i,
\hat{b}_i$, to indicate that these are different functions from
the one used in Appendix A.}.

The structure of the metric and the two-forms given in Appendix B
is almost identical to the ones for the PW flow solutions from
Section 4. The only difference is the $d\alpha_1d\alpha_2$ term in
the metric which was not present in the PW solution, however one
can show that this term does not modify the result of the null
Melvin twist so we can easily find the non-relativistic analog of
the family of fixed points found in \cite{Halmagyi:2005pn}. Since
we have gravity duals to a family of fixed points the function
$A(r)$ is simply
\begin{equation}
A(r) = \ds\frac{r}{L_{IR}}\ ,
\end{equation}
the constant $f_0$ defined in \cite{Halmagyi:2005pn} can be
written as
\begin{equation}
f_0 = \ds\frac{2^{5/3}}{3} ~.
\end{equation}
The null Melvin twist modifies the metric and the B-field and
leaves the dilaton invariant
\begin{multline}
ds^2_{10} = \hat{\Omega}^2L^2_{IR}\left(- \ds\frac{\eta^2\hat{f}_4
\hat{\Omega}^2}{L^2_{IR}}
\ds\frac{du^2}{z^4}~+~\ds\frac{1}{z^2}\left(-2dudv ~+~
d\hat{x}_1^2 +d\hat{x}_2^2 ~+~ dz^2\right)\right.
\\ \left.-
2\ds\frac{\hat{b}_1}{L^2_{IR}\cos\alpha_3}~\sigma_1\ds\frac{du}{z^2}\right)
+ds^2_{\rm int}~,\nonumber
\end{multline}
\begin{equation}
B_{(2)} = \eta\hat{\Omega}^2 ( \hat{f}_4 d\phi + \hat{f}_6
\sigma_3) \wedge \ds\frac{du}{z^2} + B_{{(2)}{\rm int}}~,
\end{equation}
where
\begin{multline}
ds^2_{\rm int} = \hat{f}_8 d\theta^2 + \hat{f}_1~ d\alpha_1^2 +
2\hat{f}_9~d\alpha_1d\alpha_2 +\hat{f}_2~d\alpha_2^2 +
\hat{f}_3~d\alpha_3^2 + \hat{f}_4~ d\phi^2 + 2\hat{f}_5 ~d\alpha_2
d\alpha_3 + 2\hat{f}_6~ d\alpha_3 d\phi + 2\hat{f}_7~ d\alpha_2
d\phi ~,
\end{multline}
and
\begin{multline}
B_{{(2)}{\rm int}} =\hat{b}_1~ d\phi \wedge d\alpha_1 + \hat{b}_2~
d\alpha_3 \wedge d\alpha_1 + \hat{b}_3~ d\alpha_2 \wedge d\alpha_1
+ \hat{b}_4~ d\theta \wedge d\alpha_2 + \hat{b}_5~ d\theta \wedge
d\alpha_1 +  \hat{b}_6~ d\phi \wedge d\alpha_2 + \hat{b}_7~
d\alpha_3 \wedge d\alpha_2 ~.
\end{multline}
Again we have performed the change of variables
\begin{equation}
u = (t+y)~, \qquad v = \ds\frac{1}{2L_{IR}^2}(t-y)~, \qquad z =
e^{-r/L_{IR}}~,\qquad \hat{x}_1 =L_{IR}^{-1}x_1~,\qquad \hat{x}_2
=L_{IR}^{-1}x_2 ~.\nonumber
\end{equation}
The RR two-form is invariant under the null Melvin twist and the
self-dual five-form transforms to
\begin{equation}
\widetilde{F}_{(5)}=F_{(5)}+\eta\left(1+\star\right)\left(\hat{\Omega}^2e^{2A}du\wedge\left(\hat{f}_4d\phi+\hat{f}_6d\alpha_3+\hat{f}_7d\alpha_2\right)\right)~.
\end{equation}

Note that the functions $\hat{f}_4$, $\hat{f}_6$, $\hat{\Omega}$
and $\ds\frac{\hat{b}_1}{\cos\alpha_3}$ depend only on $\theta$.
Their analytic form is not explicitly known for the whole family
of fixed point solutions (they are known at the KW and PW points).
However one can find numerical solutions for them
\cite{Halmagyi:2005pn}.

The construction above shows that there is a one parameters family
of supergravity solutions with Schr\"{o}dinger symmetry
interpolating between the non-relativistic cousins of the KW and
PW fixed point solutions. It should be noted also that the
function in front of the $du^2$-term in the metric, denoted by
$(-\hat{\Omega}^4\hat{f}_4)$, is negative definite ensuring the
absence spacetime pathologies for all backgrounds in the
interpolating family.

\subsection{The non-relativistic Klebanov-Witten point}

At the KW point the background simplifies significantly and we
have
\begin{equation}
\hat{b}_i = \hat{c}_i = 0~, \qquad \qquad \hat{\Omega}^2 =
f_0^{1/2}~,
\end{equation}
one finds also that
\begin{equation}
\hat{f}_4 = \ds\frac{4}{9} L_{IR}^2 f_0^{1/2}~, \qquad\qquad
\hat{f}_6 = - \ds\frac{4}{9} L_{IR}^2 f_0^{1/2} \cos\theta ~.
\end{equation}
The metric and the B-field are then
\begin{multline}
ds^2_{10} = \ds\frac{2^{5/6}}{3^{1/2}}L^2_{IR}\left(-
 \ds\frac{\eta^2}{z^4}
du^2~+~\ds\frac{1}{z^2}\left(-2dudv ~+~ d\hat{x}_1^2 +d\hat{x}_2^2
~+~ dz^2\right)\right) +ds^2_{\rm int}~,\nonumber
\end{multline}
\begin{equation}
B_{(2)} = \eta\ds\frac{2^{11/6}}{3^{3/2}}L^2_{IR}( d\phi -
\cos\theta \sigma_3) \wedge \ds\frac{du}{z^2} ~,
\end{equation}
where
\begin{multline}
ds^2_{\rm int} = \ds\frac{2^{5/6}}{3^{3/2}}L^2_{IR}\left[d\theta^2
+ \cos^2\theta \sigma_1^2+\sigma_2^2 +
\ds\frac{4\sin^2\theta}{3+\cos^2\theta}d\phi^2 +
\ds\frac{3+\cos^2\theta}{3} \left(\sigma_3 -
\ds\frac{4\cos\theta}{3+\cos^2\theta}d\phi\right)^2 \right] ~.
\end{multline}
As expected from the general calculation about the whole family of
fixed points this solution is invariant under the Schr\"{o}dinger
symmetry. Note that there is no off-diagonal $\sigma_1du$ term in
the metric and the coefficient in front of $du^2$ is a constant.
There is a further simplification in this solution since the
internal NS and RR fluxes vanish. These features are due to the
fact that the internal manifold is $T^{(1,1)}$, which is a
Sasaki-Einstein manifold and the coordinate $\phi$ used for the
null Melvin twist happens to be the Hopf fiber
\cite{Herzog:2008wg,Maldacena:2008wh,Adams:2008wt}. It is also
possible to find the finite temperature version of this solution
by putting a Schwarzschild black hole in the ${\rm AdS}_5$. Due to
the complications arising from the presence of internal fluxes and
the non-trivial dilaton the finite temperature solutions for the
rest of the family of fixed points are not known.

\section{Non-relativistic Coulomb branch RG flows}

It is natural to consider the null Melvin twist of RG flow
solutions which do not flow to a fixed point in the IR. One such
example for which the type IIB solution is explicitly known is the
RG flow solution dual to $\mathcal{N}=2^{*}$ SYM found in
\cite{Pilch:2000ue}. This geometry is dual to a particular mass
deformation of $\mathcal{N}=4$ SYM and the field theory flows to
the Coulomb branch. The background has internal NS and RR fluxes
and is thus different from the non-relativistic Coulomb branch
gravity solutions discussed in \cite{Hartnoll:2008rs}.

The background geometry, in the notation of \cite{Pilch:2000ue},
can be written as
\begin{eqnarray}
ds^2 &=& \Omega^2ds_{1,4}^2+ds_5^2\ ,\quad ds_5^2=\frac{a^2}{2}\frac{\left(cX_1X_2\right)^{1/4}}{\rho^3}\left[\frac{d\theta^2}{c}+\ds\frac{\rho^6}{4}~\cos^2\theta\left(\frac{\sigma_1^2}{cX_2}+\frac{\sigma_2^2+\sigma_3^2}{X_1}\right)+\frac{\sin^2\theta}{X_2}d\phi^2\right]\ , \nonumber\\
c &=& \cosh2\chi\ ,\quad \Omega^2=\frac{\left(cX_1X_2\right)^{1/4}}{\rho}\ , \quad X_1=\cos^2\theta+\rho^6\cosh2\chi\sin^2\theta\ , \nonumber\\
X_2 &=& \cosh2\chi\cos^2\theta+\rho^6\sin^2\theta\ ,
\end{eqnarray}
where $\sigma_i$'s are defined in (\ref{eqt: sigma}) and
$a=\sqrt{2}L_{UV}$. The metric has $SU(2)\times U(1)_{23}\times
U(1)_{\phi}$ isometry, where $U(1)_{23}$ refers to rotating
$\sigma_2$ to $\sigma_3$. As usual the dilaton-axion field can be
written as a complex scalar field $\lambda=C_{(0)}+ie^{-\Phi}$
with
\begin{eqnarray}
\lambda=i\left(\frac{1-b}{1+b}\right)\ , \quad
b=\left(\frac{c^{1/2}X_1^{1/2}-X_2^{1/2}}{c^{1/2}X_1^{1/2}+X_2^{1/2}}\right)e^{2i\phi}\
.
\end{eqnarray}
This clearly breaks the $U(1)_{\phi}$ and we are left with
$SU(2)\times U(1)_{23}$ isometry.

The background also has non-zero RR and NS forms
$C_{(2)}$,$C_{(4)}$ and $B_{(2)}$ given by
\begin{eqnarray}
\mathcal{B}_{(2)}&=&e^{i\phi}\left(\ds\frac{a_1}{2}~d\theta\wedge\sigma_1+\ds\frac{a_2}{4}~\sigma_2\wedge\sigma_3+\ds\frac{a_3}{2}~\sigma_1\wedge d\phi\right)\ ,\nonumber\\
a_1&=&-i\frac{a^2}{2}\tanh(2\chi)\cos\theta\ , \quad a_2=i\frac{a^2}{2}\frac{\rho^6\sinh(2\chi)}{X_1}\sin\theta\cos^2\theta\ , \quad a_3=\frac{a^2}{2}\frac{\sinh(2\chi)}{X_2}\sin\theta\cos^2\theta\ ,\nonumber\\
B_{(2)}&=&{\rm Re}[ \mathcal{B}_{(2)}]\ , \quad C_{(2)}={\rm Im}[ \mathcal{B}_{(2)}]\ ,\nonumber\\
C_{(4)}&=&e^{4A}\frac{X_1}{\rho^2}dt\wedge dx_1\wedge dx_2\wedge
dy\ .
\end{eqnarray}
It is clear that the two-form potentials also break the
$U(1)_{\phi}$-isometry.

The UV fixed point is obtained by setting $\chi=0$ and $\rho=1$,
where $A(r)=r/L$. To perform the null Melvin twist we choose a
$U(1)$ subgroup of the $SU(2)$ isometry along the coordinate
$\alpha_2$. The metric after the twist is given by
\begin{eqnarray}
 ds^2&=&-h\Omega^4e^{4A}\left[\eta(dt+dy)\right]^2+\Omega^2e^{2A}\left[-(dt+dy)(dt-dy)+dx_1^2+dx_2^2\right]+\Omega^2dr^2+\nonumber\\
 && 2\Omega^2e^{2A}\eta(dt+dy)\Sigma_{(1)} + ds_5^2\ ,
\end{eqnarray}
where
\begin{eqnarray}\label{eqt: he}
h&=&\frac{a^2}{8}\left(cX_1X_2\right)^{1/4}\rho^3\cos^2\theta\left[\frac{1}{cX_2}\sin^2\alpha_1\sin^2\alpha_3+\frac{1}{X_1}\left(\sin^2\alpha_1\cos^2\alpha_3+\cos^2\alpha_1\right)\right]\ ,\nonumber\\
\Sigma_{(1)}&=&\frac{a^2}{4}\sinh(2\chi)\cos\theta\left[\sin\alpha_1\sin\alpha_3\left(\frac{\sin\phi}{\cosh(2\chi)}d\theta-\frac{\cos\phi\sin\theta\cos\theta}{X_2}d\phi\right)\right. \nonumber\\
&& \left.
-\frac{\rho^6}{2X_1}\sin\theta\cos\theta\sin\phi\left(\sin\alpha_1\cos\alpha_3d\alpha_3+\sin\alpha_3\cos\alpha_1d\alpha_1\right)\right]\
.
\end{eqnarray}
The NS-NS two-form becomes
\begin{eqnarray}
\widetilde{B}_{(2)}&=&{\rm
Re}\left[\mathcal{B}_{(2)}\right]+\Omega^2e^{2A}\Gamma_{(1)}\wedge \eta \left(dt+dy\right)\ , \nonumber\\
\Gamma_{(1)}&=&\frac{a^2}{8}\left(cX_1X_2\right)^{1/4}\rho^3\cos^2\theta\left[\frac{\cos\alpha_1}{X_1}d\alpha_3+\sin\alpha_1\cos\alpha_3\sin\alpha_3\left(\frac{1}{cX_2}-\frac{1}{X_1}\right)d\alpha_1\right]+hd\alpha_2\ .\nonumber\\
\end{eqnarray}
And finally the RR forms are given by
\begin{eqnarray}
\widetilde{\lambda}&=&\lambda\ ,\nonumber\\
\widetilde{C}_{(2)}&=&{\rm Im}\left[\mathcal{B}_{(2)}\right]+C_{(0)}\Sigma_{(1)}\wedge \eta \left(dt+dy\right)\ , \nonumber\\
\widetilde{C}_{(4)}&=&C_{(4)}-\Omega^2e^{2A}C_{(2)}\wedge\Gamma_{(1)}\wedge
\eta \left(dt+dy\right)\ .
\end{eqnarray}
From the definition of $h$ in (\ref{eqt: he}), it is clear that
$h$ is a positive function\footnote{The functions $\cosh(2\chi)$
and $\rho=e^{\alpha(r)}$ are always positive for real $\chi(r)$
and real $\alpha(r)$. The family of different solutions for
$\{\chi(r), \alpha(r)\}$ represents different flows to the
$\mathcal{N}=2^{*}$ gauge theory in the parent relativistic
version. We refer to \cite{Pilch:2000ue} for more details.} except
at $\theta=\pi/2$ and hence the corresponding non-relativistic
background is again free off spacetime pathologies. At
$\theta=\pi/2$ we have $g_{uu}=0$, one can check however that the
curvature is finite at this points so nothing dramatic happens to
the ten-dimensional metric.

At the UV, the twisted background takes a pleasingly simple form
\begin{eqnarray}
ds^2&=&L^2\left(-\left(\frac{1}{4}\cos^2\theta\right)\frac{\eta^2}{z^4}du^2+\frac{1}{z^2}\left(-2du dv+d\hat{x}_1^2+d\hat{x}_2^2+dz^2\right)\right)+ L^2 ds_{UV}^2\ , \nonumber\\
ds_{UV}^2&=&d\theta^2+\frac{1}{4}\cos^2\theta\left(\sigma_1^2+\sigma_2^2+\sigma_3^2\right)+\sin^2\theta d\phi^2\ , \nonumber\\
\widetilde{B}_{(2)}&=&\frac{\eta L^2}{z^2}\left(\frac{1}{4}\cos^2\theta\right)\left(d\alpha_2+\cos\alpha_1 d\alpha_3\right)\wedge du\ ,\nonumber\\
\widetilde{C}_{(4)}&=&\frac{L^2}{z^4}dt\wedge d\hat{x}_1\wedge
d\hat{x}_2\wedge dy\ .
\end{eqnarray}
This background is invariant under the Schr\"{o}dinger symmetry
and is a special case of the general null Melvin twist of ${\rm
AdS}_5\times S^5$ discussed in Section 2 with
\begin{equation}
\eta_1=0~, \qquad\qquad \eta_2=-\eta_3\equiv\ds\frac{\eta}{2}\ .
\end{equation}
As mentioned above the $g_{uu}$ component of the metric has a zero
at $\theta=\pi/2$, however the curvature of the ten dimensional
solution is regular at this point so the background does not have
physical singularities. We want to stress again that the main
difference between the Coulomb branch solutions presented in
\cite{Hartnoll:2008rs} and the solution constructed above is the
presence of the off-diagonal terms in the metric and the
non-trivial $C_{(2)}$ and $B_{(2)}$ fluxes along the flow. The
Coulomb branch solutions of \cite{Hartnoll:2008rs} are
non-relativistic cousins of the RG flows in
\cite{Freedman:1999gk}, whereas what we have here is the
non-relativistic version of the solution in \cite{Pilch:2000ue}.

Although the $\mathcal{N}=2^{*}$ RG flow does not lead to a fixed
point, the parent relativistic flow has a rich structure in the
moduli space leading to the presence of the enhan\c{c}on locus
discussed in \cite{Evans:2000ct, Buchel:2000cn}. In the
relativistic case, this is obtained by probing the background
geometry with a probe $D3$-brane which spreads out into the
enhan\c{c}on locus. Similar computation is directly amenable for
the Melvin twisted non-relativistic background, however its dual
field theory interpretation is unclear to us at present.

\section{Non-relativistic Lunin-Maldacena solution}

The last background to which we will apply the null Melvin twist
is the Lunin-Maldacena deformation of ${\rm AdS}_5\times S^5$
\cite{Lunin:2005jy}. The background is obtained from the extremal
(or non-extremal \cite{Avramis:2007wb}) D3 brane solution by
applying S-duality, a T-duality a shift and another T-duality (TsT
transformation) on two of the $U(1)$ isometries of $S^5$ and then
one more S-duality. The resulting solution (at zero temperature)
is dual to the exactly marginal deformation of $\mathcal{N}=4$ SYM
discussed in \cite{Leigh:1995ep}. There are three $U(1)$
isometries on the $S^5$ which survive the TsT transformation so in
principle we can generate a three parameter non-relativistic
background by applying the general null Melvin twist discussed in
Section 2. We will however restrict to the case of null Melvin
twist along the Hopf fiber direction $\psi$ which corresponds to
the choice $\eta_1=\eta_2=\eta_3\equiv\eta$. The complete
untwisted Lunin-Maldacena solution at finite temperature is
presented in Appendix C, we refrain from presenting the detailed
steps of the null Melvin twist and give just the final solution.
We would like to point out that we consider the Lunin-Maldacena
solution with complex deformation parameter $\beta=\gamma -i
\sigma$, the special case of real deformation (known also as
$\gamma$-deformation) can be obtained by setting $\sigma=0$. In
the case of real deformation one does not have to apply the
S-duality and the background is obtained by a TsT transformation
of $AdS_5\times S^5$. We denote the metric and all the fields in
the solution after the twist with a tilde
\begin{eqnarray}
&&\widetilde{ds}^2_{\beta} = \mathcal{H}^{1/2}L^2 r^2 \left[ -
\ds\frac{\eta^2\mathcal{H}^{1/2}L^2 r^2 F}{1+\kappa f_4'}
(dt+dy)^2 - \ds\frac{F}{1+\kappa f_4'} dt^2 + \ds\frac{1}{1+\kappa
f_4'} dy^2 + dx_1^2 + dx_2^2 \right. \notag\\ &&\left.+
\ds\frac{dr^2}{F r^4} - \ds\frac{2\eta}{1+\kappa f_4'} \left( F dt
+ dy \right)(b_0'd\mu + b_1' d\alpha_1 + b_2' d\alpha_2 + b_3'
d\alpha_3 )\right]+ \left(f_0' + \ds\frac{\kappa b_0'^2}{1+\kappa
f_4'}\right)d\mu^2 \notag
\\&& + \left(f_1' + \ds\frac{\kappa b_1'^2}{1+\kappa f_4'}\right)d\alpha_1^2
+ \left(f_2' + \ds\frac{\kappa (b_2'^2-f_7'^2)}{1+\kappa
f_4'}\right)d\alpha_2^2+ \left(f_3' + \ds\frac{\kappa
(b_3'^2-f_6'^2)}{1+\kappa f_4'}\right)d\alpha_3^2 \\
&&+\ds\frac{f_4'}{1+ \kappa f_4'} ~d\psi^2 + \ds\frac{2f_7'}{1+
\kappa f_4'} ~d\psi d\alpha_2 + \ds\frac{2f_6'}{1+ \kappa f_4'}
~d\psi d\alpha_3 + 2 \left(f_5' + \ds\frac{\kappa
(b_2'b_3'-f_6'f_7')}{1+\kappa f_4'}\right) d\alpha_2 d\alpha_3 \notag\\
&&+ \ds\frac{2\kappa}{1+\kappa f_4'} \left[ b_0'b_1' d\mu
d\alpha_1 + b_0'b_2' d\mu d\alpha_2 + b_0'b_3' d\mu d\alpha_3 +
b_1'b_2' d\alpha_1 d\alpha_2 + b_1'b_3' d\alpha_1 d\alpha_3
\right] ~.\notag
\end{eqnarray}
The B-field has two pieces - $B_{\rm c}$ which is generated by the
Lunin-Maldacena transformation and has legs only along the compact
manifold \cite{Lunin:2005jy} and $B_{\rm nc}$ which is generated
by the null Melvin twist
\begin{equation}
\widetilde{B}_{(2)} =  B_{\text{c}}+B_{\text{nc}} ~,
\end{equation}
where
\begin{equation}
B_{\text{nc}}= - \ds\frac{\mathcal{H}^{1/2}L^2 \eta r^2}{1+\kappa
f_4'} (Fdt+dy)\wedge (f_4'd\psi + f_6' d\alpha_3 +
f_7'd\alpha_2)~,\nonumber
\end{equation}
\begin{multline}
B_{\text{c}}= \ds\frac{b_0'}{1+\kappa f_4'} ~d\psi\wedge d\mu +
\ds\frac{b_1'}{1+\kappa f_4'} ~d\psi\wedge d\alpha_1 +
\ds\frac{b_2'}{1+\kappa f_4'} ~d\psi\wedge d\alpha_2 +
\ds\frac{b_3'}{1+\kappa f_4'}~ d\psi\wedge d\alpha_3 \\\\ +
\left(b_4' + \ds\frac{\kappa b_0' f_6'}{1+\kappa f_4'} \right)
d\mu \wedge d\alpha_3 + \left(b_5' + \ds\frac{\kappa b_1'
f_6'}{1+\kappa f_4'} \right) d\alpha_1\wedge d\alpha_3 +
\left(b_6' + \ds\frac{\kappa (b_2' f_6'- b_3'f_7')}{1+\kappa f_4'}
\right) d\alpha_2 \wedge d\alpha_3 \\\\ + \ds\frac{\kappa b_0'
f_7'}{1+\kappa f_4'}~ d\mu \wedge d\alpha_2 + \ds\frac{\kappa b_1'
f_7'}{1+\kappa f_4'}~ d\alpha_1 \wedge d\alpha_2  ~.
\end{multline}
The dilaton is
\begin{equation}
\widetilde{\Phi} = \Phi - \ds\frac{1}{2}\ln ( 1+ \kappa f_4') =
\ds\frac{1}{2} \ln \left(\ds\frac{\mathcal{G}\mathcal{H}^2}{1+
\kappa f_4'}\right)~.
\end{equation}
After the null Melvin twist the RR potentials become
\begin{equation}
\widetilde{C}_{(0)} = C_{(0)} = \ds\frac{\mathcal{Q}}{\mathcal{H}}
e^{-\Phi}~,
\end{equation}
\begin{multline}
\widetilde{C}_{(2)} = C_{(2)} + \ds\frac{\mathcal{Q}}{\mathcal{H}}
e^{-\Phi}~ \ds\frac{\eta r^6 L^2 \mathcal{H}^{1/2}}{r^4+ f_4'
\eta^2 L^2 \mathcal{H}^{1/2} r_{+}^4 } \left( F dt + dy \right)
\wedge (f_4'd\chi + f_6'd\alpha_3 + f_7'd\alpha_2)~,
\end{multline}
\begin{multline}
\widetilde{C}_{(4)} = C_{(4)} + C_{(2)} \wedge \ds\frac{\eta r^6
L^2 \mathcal{H}^{1/2}}{r^4+ f_4' \eta^2 L^2 \mathcal{H}^{1/2}
r_{+}^4 } \left( F dt + dy \right) \wedge (f_4'd\chi +
f_6'd\alpha_3 + f_7'd\alpha_2) \\ + \ds\frac{\eta L^4 r^4}{4}
\left[ b_0' d\mu + b_1' d\alpha_1 + b_2' d\alpha_2 + b_3'd\alpha_3
+ 3 \sigma L^4 \left(w_1d\mu + \ds\frac{1}{2}w_2 d\alpha_1\right)
\right.\\ \left.+ \gamma L^4 \mathcal{G} \left(\ds\frac{1}{2}
\mu_1^2 (\mu_2^2 - \mu_3^2) d\alpha_3 +
\left(\ds\frac{1}{2}\mu_1^2\mu_2^2 + \ds\frac{1}{2} \mu_1^2\mu_3^2
- \mu_2^2\mu_3^2\right) d\alpha_2\right)   \right] \wedge
(dt+dy)\wedge dx_1\wedge dx_2~.
\end{multline}
In the above formulae we have defined
\begin{equation}
F(r) = 1 - \ds\frac{r_{+}^4}{r^4}~, \qquad \qquad \kappa(r) =
\mathcal{H}^{1/2} L^2\eta^2 \ds\frac{r_{+}^4}{r^2}~.
\end{equation}

For $\beta = 0 $ the background reduces to the null Melvin twist
of non-extremal D3 brane solution along the Hopf fiber found in
\cite{Herzog:2008wg,Adams:2008wt} and presented in Section 2.

By putting $r_{+}=0$ we get the non-relativistic zero temperature
Lunin-Maldacena solution. As advertised above, it is invariant
under the Schr\"{o}dinger symmetry and has dynamical exponent
$\nu=2$. The solution has non-zero NS and RR internal fluxes and
the characteristic rotation-like components of the metric
generated by the null Melvin twist.

A further generalization of the solution presented in this section
is possible. One can take the most general Lunin-Maldacena
solution with three deformation parameters
\cite{Frolov:2005dj,Imeroni:2008cr} and apply to it the general
three-parameter null Melvin twist from section 2. This will
generate a six-parameter deformation of $AdS_5\times S^5$ and the
resulting background should be invariant under the Schr\"{o}dinger
symmetry.

\section{Comments on the dual field theory}

Although we have not studied the non-relativistic theories dual to
the gravity solutions constructed above in much detail, we would
like to offer some comments.

First of all let us review the large $N_c$ gauge theories dual to
the gravity solutions before we apply the null Melvin twist. The
$\mathcal{N}=4$ SYM has three adjoint chiral superfields, denoted
by $\Phi_I$, where $I=1, 2, 3$. It is possible to consider a mass
perturbation to the superpotential of $\mathcal{N}=4$ SYM of the
form
\begin{equation}
\Delta W \sim m_1 \Phi_1^2 + m_2 \Phi_2^2~.
\end{equation}
This is a relevant operator and therefore triggers an RG flow.

The case $m_2=0$ (or $m_1=0$) was studied by Leigh and Strassler
in \cite{Leigh:1995ep} who showed that after integrating the
massive chiral superfield the theory flows to an IR fixed point
with $\mathcal{N}=1$ supersymmetry and $SU(2)\times U(1)_{R}$
global symmetry. The type IIB gravity dual of this RG flow was
constructed in \cite{Pilch:2000fu} and we applied the null Melvin
twist to it in Section 4.

The more general case when both $m_1$ and $m_2$ are non-zero and
equal to each other an $\mathcal{N}=2^*$ supersymmetric RG flow
with $SU(2)_{R}\times U(1)$ global symmetry is triggered. The
theory flows to the Coulomb branch and does not have a fixed point
in the IR \cite{Donagi:1995cf}. The gravity solution dual to this
mass deformation was found in \cite{Pilch:2000ue} and we discussed
its null Melvin twist in Section 6.

In the most general case, one can consider turning on masses for
all three chiral superfields. This theory has $\mathcal{N}=1$
supersymmetry and again undergoes an RG flow resulting in a very
rich structure in the IR \cite{Polchinski:2000uf}. However its
exact ten-dimensional gravity dual is not known so we cannot apply
the null Melvin twist to it\footnote{Linearized solutions of type
IIB were found in \cite{Polchinski:2000uf}.}.

$\mathcal{N}=4$ SYM has also a particular set of marginal
deformations parametrized by a complex parameter $\beta$ and
therefore known as the $\beta$-deformations \cite{Leigh:1995ep}.
In this case, the superpotential of $\mathcal{N}=4$ SYM is
modified in the following way
\begin{equation}
W \sim Tr(\Phi_1[\Phi_2,\Phi_3])~~~ \to ~~~ W_{\beta} \sim
Tr(e^{i\pi\beta}\Phi_1\Phi_2\Phi_3 -
e^{-i\pi\beta}\Phi_1\Phi_3\Phi_2 )~.
\end{equation}
The field theories described by this superpotential are conformal
and have $\mathcal{N}=1$ supersymmetry. Their gravity duals were
constructed in \cite{Lunin:2005jy} by a procedure very similar to
the null Melvin twist. One starts with ${\rm AdS}_5\times S^5$ and
performs an S-duality followed by a TsT transformation on two
isometry directions in the internal manifold and then one more
S-duality. This construction can be easily extended to the
non-extremal $D3$ brane solution \cite{Avramis:2007wb} which
corresponds to turning on temperature in the dual field theory. We
discussed the non-relativistic version of the Lunin-Maldacena
solution in Section 7.

Finally let us comment on the relativistic field theory dual the
family of RG flows and fixed points discussed in Section 5. We
begin with $\mathbb{Z}_2$ quiver gauge theory with an $SU(N)\times
SU(N)$ gauge group, two hypermultiplets $(A_1, B_2)$ and
$(B_1,A_2)$ and a pair of adjoint chiral superfields
$(\Phi_1,\Phi_2)$ \cite{Kachru:1998ys,Klebanov:1998hh}. The theory
is conformal and has $\mathcal{N}=2$ supersymmetry. It can be
deformed by the following  mass term
\begin{equation}
\Delta W \sim m_1 \Phi_1^2 + m_2 \Phi_2^2~.
\end{equation}
In the infrared one can integrate this mass term and find a family
of $\mathcal{N}=1$ conformal fixed points parametrized by the
ratio $m_1/m_2$. Since the $m_i$'s are in general complex numbers
this is actually a $\mathbb{CP}^{1}$ worth of conformal IR fixed
points \cite{Corrado:2004bz}. The explicit gravity duals of this
family of fixed points are known, for $m_1=m_2$ one gets the
Pilch-Warner fixed point \cite{Pilch:2000fu} and for $m_1=-m_2$
one finds the Klebanov-Witten fixed point (see Fig.1). The gravity
solutions for an arbitrary value of $m_1/m_2$ interpolating
between these two fixed points were constructed in
\cite{Halmagyi:2005pn} (see also \cite{Halmagyi:2004jy}).

\begin{figure}[t]
\centering
\includegraphics[width=5cm]{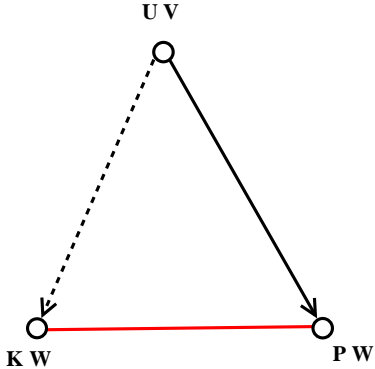}
\caption{{\it The massive RG flow from a $\mathbb{Z}_2$ orbifold
of $\mathcal{N}=4$ SYM in the UV to the KW and PW fixed points in
the IR. The horizontal (red) line represents a family of IR fixed
points interpolating between the KW and PW solutions. The solid
lines indicate that the supergravity solutions are known (at least
numerically). The structure of this RG flows and the family of
fixed points is the same after the null Melvin twist, in
particular there is a line of fixed point solutions with
Schr\"{o}dinger symmetry.}} \label{familyflow}
\end{figure}

By performing the null Melvin twist we introduce a particular
deformation of the type IIB background which is proportional to
the real parameter $\eta$. When we put $\eta=0$ in our gravity
solutions we recover the Discrete Light-Cone Quantization (DLCQ)
of the original solutions (without the null Melvin twist) along
the compact light-like direction $v\sim (t-y)$. The DLCQ amounts
to requiring all fields to be invariant under translations along
the light-like direction $v$. The way to include the deformation
introduced by the null Melvin twist was discussed in
\cite{Adams:2008wt,Bergman:2001rw}. One simply has to twist the
momentum generator $J_{v}$ by the momentum generator along the
isometry direction used for the null Melvin twist\footnote{For the
general Melvin twist of Section 2 we have $\tilde{J}_{v} = J_{v} -
\sum_{i=1}^{3}\eta_i J_{\varphi_i}$ }
\begin{equation}
\tilde{J}_{v} = J_{v} - \eta J_{\phi}~.
\end{equation}
So the deformed DLCQ that we have to perform for $\eta\neq 0$
amounts to requiring all fields to be invariant under shifts
generated by $\tilde{J}_{v}$. As noted in \cite{Adams:2008wt},
$\tilde{J}_{v}$ is not light-like for $\eta\neq 0$, therefore one
technically has a DLCQ only from the point of view of the boundary
field theory. Although our gravity solutions look quite messy the
discussion above apply to all of them which have the
Schr\"{o}dinger symmetry.

The meaning of this deformed DLCQ in the dual field theory is well
understood \cite{Adams:2008wt,Bergman:2001rw}. One has to perform
a deformed DLCQ with the twisted generator $\tilde{J}_{v}$ on the
undeformed theory dual to the gravity solution before the null
Melvin twist, now $J_{\phi}$ has to be substituted with its dual
R-symmetry generator. The momenta in the $v$ direction of all
fields get shifted by a certain amount proportional to
$\eta$\footnote{For the general null Melvin twist we have to
modify this to $P_{v} = \ds\frac{N}{R_v} + \sum_{i=1}^{3}\eta_i
Q_{i}$, where $Q_{i}$ are the R-charges under $U(1)_{\phi_i}$.}
\begin{equation}
P_{v} = \ds\frac{N}{R_v} + \eta Q\ ,
\end{equation}
where $N$ is an integer, $Q$ is the R-charge of the field under
the $U(1)$ R-symmetry that we use for the null Melvin twist and
$R_{v}$ is the radius of the compact $v$ direction, $v\sim v+2\pi
R_{v}$.

Alternatively, one can think of the field theory dual to the
gravitational solution after the null Melvin twist as a theory in
which the products of fields get modified in the following
way\footnote{These theories are called dipole theories and the
modified product is called a star product, see
\cite{Bergman:2001rw,Bergman:2000cw,Dasgupta:2000ry} for a more
detailed discussion of this kind of field theories and their
gravity duals.}
\begin{equation}
\Psi_{1} \Psi_{2} ~~~\to~~~ \Psi_{1} \star \Psi_{2} = e^{2\pi i
\eta (P_{v}^{1}Q^{2}-P_{v}^{2}Q^{1})} \Psi_{1}\Psi_{2}~,
\label{starprod}
\end{equation}
where $P_{v}^{i}$ is the momentum of the field $\Psi_{i}$ along
$v$ and $Q_{R}^{i}$ is its R-charge along the $U(1)$ direction
used for the null Melvin twist.

We can conclude that the field theory duals of the gravity
solutions that we obtained via the null Melvin twist can be
obtained by applying the deformed DLCQ procedure described above
to the field theory dual to the gravity solution before the null
Melvin twist. In particular the solution discussed in Section 4 is
dual to the deformed DLCQ of a mass deformation of $\mathcal{N}=4$
SYM where one gives mass to one of the chiral superfields. We find
that the gravity dual suggests, much like its parent relativistic
theory, that the non-relativistic theory flows to an IR fixed
point after the massive field is integrated out, at this fixed
point the field theory has dynamical exponent $\nu=2$ and is
invariant under the Schr\"{o}dinger symmetry. The solution in
Section 7 is dual to a deformed DLCQ of the $\beta$-deformation of
$\mathcal{N}=4$ SYM discussed in \cite{Leigh:1995ep,Lunin:2005jy},
the non-relativistic field theory is again Schr\"{o}dinger
invariant. The solution of Section~6 is a bit more subtle since it
does not flow to an IR fixed point. One can think of it in the
following way - perform a deformed DLCQ of $\mathcal{N}=4$ SYM and
then give equal masses to two of the chiral superfields. The
theory undergoes an RG flow similar to the RG flow of its parent
relativistic theory, there is no IR fixed point and the theory
flows to a Coulomb branch. Finally the solutions in Section 5
should be dual to a family of non-relativistic IR fixed points
with Schr\"{o}dinger symmetry, which can be obtained by a mass
deformation of the deformed DLCQ of the $\mathcal{N}=2$ quiver
gauge theory obtained by a $\mathbb{Z}_2$ orbifold of
$\mathcal{N}=4$ SYM. Note that in all cases we considered we have
at least $U(1)$ R-symmetry and we used exactly the corresponding
isometry direction in the internal manifold for the null Melvin
twist. However in all cases the $U(1)$ R-symmetry is a different
$U(1)$ subgroup of the $SO(6)$ R-symmetry of $\mathcal{N}=4$ SYM
which will lead to a different definition of the star product
(\ref{starprod}). Note also that since we are discussing
supergravity solutions, strictly speaking, they should be
considered as duals to field theories at large $N_c$.

Let us also comment on the meaning of the parameter $\eta$. As
pointed out in refs. \cite{Herzog:2008wg,Adams:2008wt}, the
particle number density in the non-relativistic field theory is
proportional to some power of~$\eta/R_{v}$. However for the
extremal gravity solutions (i.e. $T=0$) this parameter can be set
to unity by rescaling $u\to \eta u$, $v\to \eta^{-1} v$. Since the
RG flow solutions considered here are only known for $T=0$ we can
remove the dependence of the metric on $\eta$, however we chose to
keep $\eta$ explicit to emphasize that it has a physical meaning
and cannot be removed at finite temperature. We found the finite
temperature non-relativistic Lunin-Maldacena solution and thus the
dual field theory has two important parameters: the temperature
$T$ and the particle number density which is related to $\eta$. It
will be interesting to study the thermodynamics of this theory and
see if the marginal deformation parameter $\beta$ affects any of
the interesting thermodynamical quantities. For this finite
temperature solution we can of course take the limit $\eta \to 0~,
T=\text{fixed}$ which will lead to the trivial case of zero
particle number density in the dual field theory.

\section{Conclusions}

In this note we found the non-relativistic generalizations of some
known gravity solutions which are dual to RG flows and marginal
deformations of $\mathcal{N}=4$ SYM. We found that when the
original type IIB background dual to the relativistic theory has
an ${\rm AdS}_5$ fixed point the non-relativistic geometry
obtained after the null Melvin twist has the Schr\"{o}dinger
symmetry. This is quite a generic feature and it should persist
for other ${\rm AdS}$ compactifications with internal $U(1)$
isometry to which one can apply the null Melvin twist.

Another general feature of all backgrounds that we discussed is
the existence of RR and NS internal fluxes. In particular after
the null Melvin twist the internal NS flux translates into
off-diagonal, rotation-like, terms in the metric. More generally
it will be interesting to understand if there is a consistent
truncation of Schr\"{o}dinger invariant ten dimensional IIB
backgrounds with internal fluxes to five dimensions  and if one
can realize holographic RG flows directly in the five-dimensional
gravitational theory.

It will be also quite interesting to look at the
Polchinski-Strassler solution \cite{Polchinski:2000uf} and see how
much of its rich structure is present in its non-relativistic
version. The exact gravity solution dual to this theory is not
known so one cannot straightforwardly apply the null Melvin twist.
There are certainly other known solutions of type IIB and
eleven-dimensional supergravity which are dual to RG flows of
relativistic gauge theories \cite{Corrado:2001nv,othersolns}. It
will be interesting to find their non-relativistic generalizations
and understand the dual field theories. In particular the
eleven-dimensional solution of \cite{Corrado:2001nv} realizes a
holographic RG flow between $AdS_4\times S^7$ in the UV and an
$AdS_4$ compactification with fluxes in the IR, which is the
eleven-dimensional analog of the solution in \cite{Pilch:2000fu}.
It should be possible to find the non-relativistic version of this
solution which should be dual to a non-relativistic conformal
field theory in $1+1$ dimensions.

Since one of the goals of extending gauge/gravity duality to
non-relativistic field theories is to be able to gain some
understanding of these theories at strong coupling and finite
temperature, it will be quite interesting to construct the finite
temperature cousins of the non-relativistic RG flows that were
discussed here\footnote{ See \cite{Buchel:2003ah} for some work on
the $\mathcal{N}=2^{*}$ PW RG flow at finite temperature.}. This
will be a rather non-trivial task and maybe one has to look for
approximate solutions and extract the interesting physics from
them.

The general null Melvin twist discussed in Section 2 relies on the
$U(1)\times U(1) \times U(1)$ subgroup of the $SO(6)$ isometry
group of $S^5$. One can use this to generate a large class of
non-relativistic type IIB backgrounds by applying the null Melvin
twist to solutions of the form ${\rm AdS}_5\times L^{p,q,r}$ where
$L^{p,q,r}$ are the five-dimensional Sasaki-Einstein manifolds
constructed in \cite{Cvetic:2005ft}. These manifolds have
$U(1)\times U(1) \times U(1)$ isometry and are a generalization of
the $Y^{p,q}$ Sasaki-Einstein manifolds found in
\cite{Gauntlett:2004yd}. For generic values of
$(\eta_1,\eta_2,\eta_3)$ we expect that these non-relativistic
solutions will break supersymmetry completely, but there might be
special choices of $(\eta_1,\eta_2,\eta_3)$ for which some Killing
spinors are preserved \cite{progress}.

This is a modest first step in constructing holographic duals of
non-relativistic CFT's deformed by relevant or marginal operators,
however it can be a hint for the construction of a more general
ansatz for other gravity solutions dual to non-relativistic CFT's.
These may include solutions with different amounts of
supersymmetry \cite{progress}, as well as solutions with different
dynamical exponents and brane wave deformations
\cite{Hartnoll:2008rs}. We hope to return to some of these
problems in the near future.

\bigskip
\bigskip
\textbf{\large Acknowledgments}
\bigskip

We would like to thank Nick Halmagyi, Mike Mulligan and Kentaroh
Yoshida for useful discussions and comments. We are grateful to
Mukund Rangamani for numerous illuminating conversations and
explanations. We would also like to thank Clifford Johnson,
Krzysztof Pilch and Nick Warner for their interest in this work,
many useful discussions and for the constant encouragement. We
would especially like to thank Clifford Johnson for valuable
comments on the manuscript. This work is supported in part by the
DOE grant DE-FG03-84ER-40168. NB is supported also by a Graduate
Fellowship from KITP and in part by the National Science
Foundation under Grant No. PHY05-51164.

\renewcommand{\theequation}{A.\arabic{equation}}
\setcounter{equation}{0}  
\section*{Appendix A. Null Melvin twist of the Pilch-Warner flow}
\addcontentsline{toc}{section}{Appendix A. Null Melvin twist of the Pilch-Warner flow}

Here we present in detail the null Melvin twist of the
$\mathcal{N}=1$ PW RG flow solution. It is convenient to rewrite
the solution in a more compact form. The metric is (we put
$y\equiv x_3$)
\begin{multline}
ds^2_{10} = \Omega^2 e^{2A} ( - dt^2 + dy^2 + dx_1^2 +dx_2^2) +
\Omega^2 dr^2 + \ds\frac{L^2 \Omega^2}{\rho^2\cosh^2\chi}
d\theta^2 \\\\+ f_1~ d\alpha_1^2 + f_2~d\alpha_2^2 +
f_3~d\alpha_3^2 + f_4~ d\phi^2 + 2f_5 ~d\alpha_2 d\alpha_3 + 2f_6~
d\alpha_3 d\phi + 2f_7~ d\alpha_2 d\phi  ~.
\end{multline}
The NS and RR two-form potentials are
\begin{multline}
B_{(2)} = b_1~ d\phi \wedge d\alpha_1 + b_2~ d\alpha_3 \wedge
d\alpha_1 + b_3~ d\alpha_2 \wedge d\alpha_1 + b_4~ d\theta \wedge
d\alpha_2 + b_5~ d\theta \wedge d\alpha_1 \\\\ +  b_6~ d\phi
\wedge d\alpha_2 +  b_7~ d\alpha_3 \wedge d\alpha_2 ~.
\end{multline}
\begin{multline}
C_{(2)} = c_1~ d\phi \wedge d\alpha_1 + c_2~ d\alpha_3 \wedge
d\alpha_1 + c_3~ d\alpha_2 \wedge d\alpha_1 + c_4~ d\theta \wedge
d\alpha_2 + c_5~ d\theta \wedge d\alpha_1 \\\\ +  c_6~ d\phi
\wedge d\alpha_2 + + c_7~ d\alpha_3 \wedge d\alpha_2 ~.
\end{multline}
Where we have defined
\begin{eqnarray}
f_1 &=& \ds\frac{L^2\rho^2}{\Omega^2}
\ds\frac{\cos^2\theta}{4}~,\notag\\
f_2 &=& \ds\frac{L^2\rho^2}{\Omega^2}
\ds\frac{\cos^2\theta}{4} \sin^2\alpha_1 + L^2\Omega^2 \cos^2\alpha_1\left[ \ds\frac{\rho^4}{X_1^2}\ds\frac{\cos^4\theta}{4} + \ds\frac{\sin^2\theta\cos^2\theta}{4\rho^2\cosh^2\chi} \left(1- \ds\frac{1-\rho^6}{X_1}\cos^2\theta\right)^2 \right]~,\notag\\
f_3 &=& L^2\Omega^2 \left[ \ds\frac{\rho^4}{X_1^2}\ds\frac{\cos^4\theta}{4} + \ds\frac{\sin^2\theta\cos^2\theta}{4\rho^2\cosh^2\chi} \left(1- \ds\frac{1-\rho^6}{X_1}\cos^2\theta\right)^2 \right]~,\\
f_4 &=& \ds\frac{L^2\Omega^2\rho^4}{X_1^2}\left(1 - \ds\frac{3}{2}\cos^2\theta\right)^2 + \ds\frac{L^2 \Omega^2}{\rho^2\cosh^2\chi} \sin^2\theta\cos^2\theta \left(\ds\frac{3}{2} + \ds\frac{1-\rho^6}{X_1}\left(1 - \ds\frac{3}{2}\cos^2\theta\right)\right)^2~,\notag\\
f_5 &=& f_3 \cos\alpha_1~, \notag\\
f_6 &=& \ds\frac{L^2\Omega^2\rho^4 }{X_1^2} \ds\frac{\cos^2\theta}{2}\left(1 - \ds\frac{3}{2}\cos^2\theta\right) \notag\\ &-& \ds\frac{L^2 \Omega^2}{\rho^2\cosh^2\chi} \ds\frac{\sin^2\theta\cos^2\theta}{2} \left(1 - \ds\frac{1-\rho^6}{X_1}\cos^2\theta\right) \left(\ds\frac{3}{2} + \ds\frac{1-\rho^6}{X_1}\left(\sin^2\theta - \ds\frac{1}{2}\cos^2\theta\right)\right)~,\notag\\
f_7 &=& f_6 \cos\alpha_1~,\notag
\end{eqnarray}
and
\begin{eqnarray}
b_1 &=& \ds\frac{L^2}{4} \tanh\chi \cos^2\theta\sin\theta\cos\alpha_3 \left(\ds\frac{3}{2} + \ds\frac{1-\rho^6}{X_1}\left(1 - \ds\frac{3}{2}\cos^2\theta\right)\right) ~,\notag\\
b_2 &=& \ds\frac{L^2}{4} \tanh\chi
\cos^2\theta\sin\theta\cos\alpha_3
\left(\ds\frac{1-\rho^6}{2X_1}\cos^2\theta-\ds\frac{1}{2}\right)~,
\qquad\qquad b_3 = b_2 \cos\alpha_1~,\\
b_4 &=& - \ds\frac{L^2}{4} \tanh\chi \cos\theta
\cos\alpha_3\sin\alpha_1~, \qquad\qquad b_5 = \ds\frac{L^2}{4}
\tanh\chi
\cos\theta \sin\alpha_3~,\notag\\
b_6 &=& b_1 ~ \tan\alpha_3 \sin\alpha_1~,\qquad\qquad b_7 = b_2 ~
\tan\alpha_3 \sin\alpha_1~,\notag
\end{eqnarray}
and
\begin{eqnarray}
c_1 &=& - b_6~\ds\frac{1}{\sin\alpha_1} = - b_1 ~ \tan\alpha_3~,
\qquad\qquad c_2 = -b_7~
\ds\frac{1}{\sin\alpha_1} = -b_2 ~ \tan\alpha_3~,\notag\\\notag\\
c_3 &=& - b_7~\cot\alpha_1 = -b_3 ~ \tan\alpha_3~,
\qquad\qquad c_4 = -b_4 ~ \tan\alpha_3~,\\\notag\\
c_5 &=& - b_4~\ds\frac{1}{\sin\alpha_1} = b_5 \cot\alpha_3~,
\qquad c_6 = b_1~ \sin\alpha_1 = b_6 \cot\alpha_3~, \qquad c_7 =
b_2~\sin\alpha_1 = b_7 \cot\alpha_3 ~.\notag
\end{eqnarray}
Now we are ready to the apply the five steps of the null Melvin
twist.

\bigskip

\textit{Step 1} Perform a boost in the $(t,y)$ plane with a
parameter $\gamma_0$
\begin{equation}
t \to ct - sy ~, \qquad\qquad y \to -st + cy~,
\end{equation}
where
\begin{equation}
c = \cosh \gamma_0 ~, \qquad\qquad s = \sinh\gamma_0~,
\end{equation}
note that $c^2-s^2 = 1$. The whole background is invariant under
this boost so nothing is changed.

\bigskip

\textit{Step 2} Perform T-duality along $y$. To avoid clutter we
will not show the explicit changes in the RR-forms at each steps,
but present the final result.

At this step the B-field is invariant under this T-duality, the
only changes are in the metric and the dilaton which take the form
\begin{multline}
ds^2_{10} = \ds\frac{1}{\Omega^2 e^{2A}}dy^2 +  \Omega^2 e^{2A} (
- dt^2 + dx_1^2 +dx_2^2) + \Omega^2 dr^2 + \ds\frac{L^2
\Omega^2}{\rho^2\cosh^2\chi} d\theta^2 \\\\+ f_1~ d\alpha_1^2 +
f_2~d\alpha_2^2 + f_3~d\alpha_3^2 + f_4~ d\phi^2 + 2f_5 ~d\alpha_2
d\alpha_3 + 2f_6~ d\alpha_3 d\phi + 2f_7~ d\alpha_2 d\phi ~,
\end{multline}
\begin{equation}
\widetilde{\Phi} = \Phi - \log ( \Omega e^{A} )~.
\end{equation}

\bigskip

\textit{Step 3} Perform a shift $\phi \to \phi + a y$. The dilaton
is invariant under this shift, but the metric and the B-field
change to
\begin{multline}
ds^2_{10} = \left(\ds\frac{1}{\Omega^2 e^{2A}} + a^2
f_4\right)dy^2 + 2af_4 dyd\phi + 2af_6 dyd\alpha_3 +2af_7
dyd\alpha_2+ \Omega^2 e^{2A} ( - dt^2 + dx_1^2 +dx_2^2) + \Omega^2
dr^2\\\\ + \ds\frac{L^2 \Omega^2}{\rho^2\cosh^2\chi} d\theta^2 +
f_1~ d\alpha_1^2 + f_2~d\alpha_2^2 + f_3~d\alpha_3^2 + f_4~
d\phi^2 + 2f_5 ~d\alpha_2 d\alpha_3 + 2f_6~ d\alpha_3 d\phi +
2f_7~ d\alpha_2 d\phi  ~,
\end{multline}
\begin{multline}
B_{(2)} = a b_1~ dy\wedge d\alpha_1 + a b_6~ dy\wedge d\alpha_2 +
b_1~ d\phi \wedge d\alpha_1 + b_2~ d\alpha_3 \wedge d\alpha_1 +
b_3~ d\alpha_2 \wedge d\alpha_1 \\\\+ b_4~ d\theta \wedge
d\alpha_2 + b_5~ d\theta \wedge d\alpha_1  +  b_6~ d\phi \wedge
d\alpha_2 +  b_7~ d\alpha_3 \wedge d\alpha_2 ~.
\end{multline}

\bigskip

\textit{Step 4} Perform one more T-duality along $y$. All NS
fields change (not all components change though) under this
transformation and the end result is
\begin{multline}
ds^2_{10} = h_1 dy^2 - 2ab_1 h_1 dyd\alpha_1 - 2ab_6 h_1
dyd\alpha_2+ \Omega^2 e^{2A} ( - dt^2 + dx_1^2 +dx_2^2) + \Omega^2
dr^2\\\\ + \ds\frac{L^2 \Omega^2}{\rho^2\cosh^2\chi} d\theta^2 +
(f_1+a^2b_1^2h_1)~ d\alpha_1^2 + (f_2+a^2
(b_6^2-f_7^2)h_1)~d\alpha_2^2 + (f_3 - a^2f_6^2h_1)~d\alpha_3^2 +
\ds\frac{f_4}{1+ a^2 f_4 \Omega^2e^{2A}}~ d\phi^2 \\\\+
2a^2b_1b_6h_1 d\alpha_1d\alpha_2 + 2(f_5-a^2 f_6f_7h_1) ~d\alpha_2
d\alpha_3 + \ds\frac{2f_6}{1+ a^2 f_4 \Omega^2e^{2A}}~ d\alpha_3
d\phi + \ds\frac{2f_7}{1+ a^2 f_4 \Omega^2e^{2A}}~ d\alpha_2 d\phi
~,
\end{multline}
\begin{multline}
B_{(2)} = a f_4h_1~ d\phi\wedge dy + a f_7h_1~ d\alpha_2\wedge dy
+ a f_6h_1~ d\alpha_3\wedge dy \\\\+ (b_1 - a^2h_1 b_1f_4)~ d\phi
\wedge d\alpha_1 + (b_2 - a^2h_1 b_1f_6)~ d\alpha_3 \wedge
d\alpha_1 + (b_3 - a^2h_1 b_1f_7)~ d\alpha_2 \wedge d\alpha_1
\\\\+ b_4~ d\theta \wedge d\alpha_2 + b_5~ d\theta \wedge
d\alpha_1  +  (b_6 - a^2h_1 b_6f_4)~ d\phi \wedge d\alpha_2 + (b_7
- a^2h_1 b_6f_6)~ d\alpha_3 \wedge d\alpha_2 ~.
\end{multline}
\begin{equation}
\widetilde{\Phi} = \Phi - \log\left(1 + a^2 f_4 \Omega^2
e^{2A}\right)~,
\end{equation}

where we have defined

\begin{equation}
h_1 = \ds\frac{\Omega^2 e^{2A}}{1 + a^2 f_4 \Omega^2 e^{2A}}~,
\end{equation}

\bigskip

\textit{Step 5} Perform one more boost in the $(t,y)$ plane with a
parameter $-\gamma_0$
\begin{equation}
t \to ct + sy ~, \qquad\qquad y \to st + cy~,
\end{equation}
then take the following limit
\begin{equation}
a \to 0  ~, \qquad\qquad \gamma_0 \to \infty~, \qquad\qquad as =
ac = \eta~.
\end{equation}
The final form of the metric is
\begin{multline}
ds^2_{10} = - f_4\Omega^4e^{4A} [\eta (dt+dy)]^2 - \Omega^2
e^{2A}(dt-dy)(dt+dy) - 2 b_1 \Omega^2 e^{2A} [\eta
(dt+dy)]d\alpha_1 - 2 b_6 \Omega^2 e^{2A} [\eta (dt+dy)]d\alpha_2
\\\\+  \Omega^2 e^{2A} (dx_1^2 +dx_2^2) + \Omega^2 dr^2 +
\ds\frac{L^2 \Omega^2}{\rho^2\cosh^2\chi} d\theta^2 \\\\+ f_1~
d\alpha_1^2 + f_2~d\alpha_2^2 + f_3~d\alpha_3^2 + f_4~ d\phi^2 +
2f_5 ~d\alpha_2 d\alpha_3 + 2f_6~ d\alpha_3 d\phi + 2f_7~
d\alpha_2 d\phi  ~.
\end{multline}
The B-field and the dilaton are
\begin{multline}
B_{(2)} = f_4\Omega^2 e^{2A} d\phi\wedge [\eta (dt+dy)] +
f_7\Omega^2 e^{2A} d\alpha_2\wedge [\eta (dt+dy)] + f_6\Omega^2
e^{2A} d\alpha_3\wedge [\eta (dt+dy)]\\\\+b_1~ d\phi \wedge
d\alpha_1 + b_2~ d\alpha_3 \wedge d\alpha_1 + b_3~ d\alpha_2
\wedge d\alpha_1 + b_4~ d\theta \wedge d\alpha_2 + b_5~ d\theta
\wedge d\alpha_1 +  b_6~ d\phi \wedge d\alpha_2 + b_7~ d\alpha_3
\wedge d\alpha_2 ~,\nonumber
\end{multline}
\begin{equation}
\widetilde{\Phi} = \Phi - \log\left(1 + a^2 f_4 \Omega^2
e^{2A}\right) \to \Phi~.
\end{equation}
Note that the metric on the compact manifold is the same after the
Melvin twist. Note also that there are off-diagonal metric
coefficients $g_{u\alpha_1}$ and $g_{u\alpha_2}$ where $u =(t+y)$.
These metric coefficients vanish at the UV fixed point because
$b_1\sim b_6 \sim \sinh\chi$ and $\chi = 0$. However at the IR
fixed point the off-diagonal metric coefficients are not
vanishing.

In the RR-sector, $G_{(3)}=dC_{(2)}$ remains invariant, but
$F_{(5)}$ changes under the Melvin twist procedure. Here we
present the final form of the five-form flux
\begin{multline}
\widetilde{F}_{(5)}=(1+\star) [dx^0\wedge dx^1\wedge dx^2\wedge dy
\wedge (w_r(r,\theta)dr + w_{\theta}(r,\theta) d\theta)\\+
\left(\Omega^2e^{2A}\right)du\wedge\left(f_4d\phi+f_6d\alpha_3+f_7d\alpha_2\right)\wedge
G_{(3)}]~.
\end{multline}
%

\renewcommand{\theequation}{B.\arabic{equation}}
\setcounter{equation}{0}  
\section*{Appendix B. Family of fixed points}
\addcontentsline{toc}{section}{Appendix B. Family of fixed points}

This Appendix is devoted to a short review of the solution found
in \cite{Halmagyi:2005pn}. The metric is (note that the $\sigma_i$
in \cite{Halmagyi:2005pn} are the same as our $\sigma_i$)
\begin{multline}
ds^2_{10} = \hat{\Omega}^2 e^{2A} ( - dt^2 + dy^2 + dx_1^2
+dx_2^2) + \hat{\Omega}^2 dr^2 + \hat{f}_8 d\theta^2 \\\\+
\hat{f}_1~ d\alpha_1^2 + 2\hat{f}_9~d\alpha_1d\alpha_2
+\hat{f}_2~d\alpha_2^2 + \hat{f}_3~d\alpha_3^2 + \hat{f}_4~
d\phi^2 + 2\hat{f}_5 ~d\alpha_2 d\alpha_3 + 2\hat{f}_6~ d\alpha_3
d\phi + 2\hat{f}_7~ d\alpha_2 d\phi ~.
\end{multline}
The NS and RR two-form potentials are
\begin{multline}
B_{(2)} = \hat{b}_1~ d\phi \wedge d\alpha_1 + \hat{b}_2~ d\alpha_3
\wedge d\alpha_1 + \hat{b}_3~ d\alpha_2 \wedge d\alpha_1 +
\hat{b}_4~ d\theta \wedge d\alpha_2 + \hat{b}_5~ d\theta \wedge
d\alpha_1 \\\\ +  \hat{b}_6~ d\phi \wedge d\alpha_2 +  \hat{b}_7~
d\alpha_3 \wedge d\alpha_2 ~,
\end{multline}
\begin{multline}
C_{(2)} = \hat{c}_1~ d\phi \wedge d\alpha_1 + \hat{c}_2~ d\alpha_3
\wedge d\alpha_1 + \hat{c}_3~ d\alpha_2 \wedge d\alpha_1 +
\hat{c}_4~ d\theta \wedge d\alpha_2 + \hat{c}_5~ d\theta \wedge
d\alpha_1 \\\\ +  \hat{c}_6~ d\phi \wedge d\alpha_2 + + \hat{c}_7~
d\alpha_3 \wedge d\alpha_2 ~.
\end{multline}
There is also a non-trivial dilaton, $\Phi(\theta)$, for the
family of fixed points. It vanishes at the PW and KW fixed points.
Above we have defined
\begin{eqnarray}
\hat{f}_1 &=& L^2_{IR}\hat{\Omega}^{-2} (A_1^2 \cos^2\alpha_3+
A_2^2 \sin^2\alpha_3)~, \qquad \hat{f}_2 =
L^2_{IR}\hat{\Omega}^{-2}\left[ (A_1^2 \sin^2\alpha_3+ A_2^2
\cos^2\alpha_3)\sin^2\alpha_1 + A_3^2 \cos^2\alpha_1
\right]~,\notag\\\notag\\
\hat{f}_3 &=& L^2_{IR}\hat{\Omega}^{-2} A_3^2~, \qquad\qquad
\hat{f}_4 =
L^2_{IR}\hat{\Omega}^{-2} (A_5^2+A_3^2B_1^2)~, \qquad\qquad \hat{f}_5 = \hat{f}_3 \cos\alpha_1~, \notag\\\\
\hat{f}_6 &=& L^2_{IR}\hat{\Omega}^{-2} A_3^2 B_1~, \qquad\qquad \hat{f}_7 = \hat{f}_6\cos\alpha_1~,\notag\\\notag\\
\hat{f}_8 &=& L^2_{IR}\hat{\Omega}^{-2} A_4^2 ~, \qquad\qquad
\hat{f}_9= L^2_{IR}\hat{\Omega}^{-2} (A_1^2-A_2^2)\sin\alpha_3
\cos\alpha_3\sin\alpha_1~,\\ \nonumber\\
\hat{\Omega}^2&=&-\frac{3}{2}A_3A_4A_5\frac{1}{\left(A_1A_2\right)'}~,\notag
\end{eqnarray}
where $'$ denotes taking derivative with respect to $\theta$ and
\begin{eqnarray}
\hat{b}_1 &=& - L^2_{IR}F_2 \cos\alpha_3~, \qquad \hat{b}_2 = -
L^2_{IR}F_1 \cos\alpha_3~, \qquad \hat{b}_3 = \hat{b}_2
\cos\alpha_1~, \qquad \hat{b}_4 =  L^2_{IR}F_3 \cos\alpha_3 \sin\alpha_1~,\notag\\\\
\hat{b}_5 &=& - L^2_{IR}F_3 \sin\alpha_3~, \qquad \hat{b}_6 =
\hat{b}_1 \tan\alpha_3\sin\alpha_1~, \qquad \hat{b}_7 = \hat{b}_2
\tan\alpha_3\sin\alpha_1~.\notag
\end{eqnarray}
Also
\begin{eqnarray}
\hat{c}_1 &=& - L^2_{IR}H_2 \sin\alpha_3~, \qquad \hat{c}_2 = -
L^2_{IR}H_1 \sin\alpha_3~, \qquad \hat{c}_3 = \hat{c}_2
\cos\alpha_1~, \qquad \hat{c}_4 = - L^2_{IR}H_3 \sin\alpha_3 \sin\alpha_1~,\notag\\\\
\hat{c}_5 &=& - L^2_{IR}H_3 \cos\alpha_3~, \qquad \hat{c}_6 =
-\hat{c}_1 \cot\alpha_3\sin\alpha_1~,\qquad \hat{c}_7 = \hat{c}_2
\cot\alpha_3\sin\alpha_1~.\notag
\end{eqnarray}
The functions $A_i$ and $B_1$ depend only ont $\theta$ and are
defined in \cite{Halmagyi:2005pn}. We have introduced six new
functions of $\theta$ -- $F_i$ and $H_i$. They are related to the
functions used in \cite{Halmagyi:2005pn} as follows. The
three-from fluxes in \cite{Halmagyi:2005pn} are
\begin{multline}
H_{(3)} = dB_{(2)} = (g_1+g_4)
\ds\frac{A_1A_3A_4}{\hat{\Omega}^3}~ \sigma_1\wedge\sigma_3 \wedge
d\theta + \ds\frac{A_1A_4}{\hat{\Omega}^3}\left[(g_1+g_4)A_3B_1 -
(g_2+g_5)A_5 \right]\sigma_1\wedge d\phi \wedge d\theta \\
+ (-ig_3+ig_6) \ds\frac{A_2A_3A_5}{\hat{\Omega}^3}~
\sigma_2\wedge\sigma_3 \wedge d\phi ~,
\end{multline}
\begin{multline}
G_{(3)} = dC_{(2)} = (g_4-g_1)
\ds\frac{A_2A_3A_4}{\hat{\Omega}^3}~ \sigma_2\wedge\sigma_3 \wedge
d\theta + \ds\frac{A_2A_4}{\hat{\Omega}^3}\left[(g_4-g_1)A_3B_1 -
(g_5-g_2)A_5 \right]\sigma_2\wedge d\phi \wedge d\theta \\
+ (-ig_3-ig_6) \ds\frac{A_1A_3A_5}{\hat{\Omega}^3}~
\sigma_1\wedge\sigma_3 \wedge d\phi ~.
\end{multline}
One can show that these fluxes come from the following potentials
\begin{equation}
B_{(2)} = F_1(\theta)~ \sigma_1\wedge\sigma_3 + F_2(\theta)
\sigma_1\wedge d\phi + F_3(\theta) \sigma_2\wedge d\theta ~,
\end{equation}
\begin{equation}
C_{(2)} = H_1(\theta)~ \sigma_2\wedge\sigma_3 + H_2(\theta)
\sigma_2\wedge d\phi + H_3(\theta) \sigma_1\wedge d\theta~,
\end{equation}
if the following identities hold (here $'=\frac{d}{d\theta}$ )
\begin{eqnarray}
F_1' - F_3 &=& (g_1+g_4) \ds\frac{A_1A_3A_4}{\hat{\Omega}^3}~,
\qquad  F_2' =
\ds\frac{A_1A_4}{\hat{\Omega}^3}\left[(g_1+g_4)A_3B_1 -
(g_2+g_5)A_5
\right]~,\\
F_2 &=& (-ig_3+ig_6) \ds\frac{A_2A_3A_5}{\hat{\Omega}^3}~.\notag
\end{eqnarray}
\begin{eqnarray}
H_1' + H_3 &=& (g_4-g_1) \ds\frac{A_2A_3A_4}{\hat{\Omega}^3}~,
\qquad  H_2' =
\ds\frac{A_2A_4}{\hat{\Omega}^3}\left[(g_4-g_1)A_3B_1 -
(g_5-g_2)A_5
\right]~,\\
H_2 &=& (ig_3+ig_6) \ds\frac{A_1A_3A_5}{\hat{\Omega}^3}~.\notag
\end{eqnarray}
There is also the usual self-dual five-form flux
\begin{equation}
F_{(5)}=f_0 dt\wedge dx^1\wedge dx^2\wedge dy\wedge dr+
f_0\hat{\Omega}^{-10}\left(A_1A_2A_3A_4A_5\right)\sigma_1\wedge\sigma_2\wedge\left(\sigma_3+B_1d\phi+B_2d\theta\right)\wedge
d\theta\wedge d\phi\ .
\end{equation}
%

\renewcommand{\theequation}{C.\arabic{equation}}
\setcounter{equation}{0}  
\section*{Appendix C. Lunin-Maldacena at finite temperature}
\addcontentsline{toc}{section}{Appendix C. Lunin-Maldacena at finite temperature}

Here we will review the finite temperature Lunin-Maldacena
solution \cite{Lunin:2005jy} found in \cite{Avramis:2007wb}. The
metric is
\begin{multline}
ds^2_{\beta} = \mathcal{H}^{1/2}L^2 \left[ -r^2
\left(1-\ds\frac{r_{+}^4}{r^4}\right)dt^2 + r^2 (dx_1^2+dx_2^2
+dy^2) +
\left(1-\ds\frac{r_{+}^4}{r^4}\right)^{-1}\ds\frac{dr^2}{r^2}
\right] \\\\+ \mathcal{H}^{1/2}L^2 \left[ \ds\sum_{i=1}^3 (
d\mu_i^2 + \mathcal{G}\mu_i^2d\varphi_i^2 ) +
\mathcal{G}|\beta|^2L^4 \mu_1^2\mu_2^2\mu_3^2 (d\varphi_1+
d\varphi_2 + d\varphi_3)^2 \right].
\end{multline}
The NS and RR fields in the deformed solution are (we have set the
dilaton in the undeformed ${\rm AdS}_5\times S^5$ solution to
zero)
\begin{equation}
B_{(2)} = \gamma \mathcal{G} \mathcal{B}_2 -\sigma\mathcal{A}_2~,
\qquad\qquad e^{2\Phi} = \mathcal{G}\mathcal{H}^2~.
\end{equation}
\begin{equation}
C_{(0)} = \mathcal{H}^{-1} \mathcal{Q}~, \qquad\qquad C_{(2)} =
-\gamma \mathcal{A}_{2} - \sigma \mathcal{G}\mathcal{B}_2 ~.
\end{equation}
\begin{equation}
C_{(4)} = \mathcal{A}_4 - \gamma^2 \mathcal{G}\mathcal{B}_2 \wedge
\mathcal{A}_2 + \ds\frac{1}{2} \gamma\sigma \mathcal{A}_2 \wedge
\mathcal{A}_2~, \qquad\qquad C_{(6)} = B_{(2)}\wedge C_{(4)} ~.
\end{equation}
Where
\begin{equation}
\beta = \gamma - i\sigma ~, \qquad \mathcal{Q} = L^4 \gamma\sigma
( \mu_1^2\mu_2^2 + \mu_1^2\mu_3^2 + \mu_2^2\mu_3^2)~,
\end{equation}
\begin{equation}
\mathcal{G} = \ds\frac{1}{1+L^4|\beta|^2 ( \mu_1^2\mu_2^2 +
\mu_1^2\mu_3^2 + \mu_2^2\mu_3^2)}~, \qquad \mathcal{H} =
1+L^4\sigma^2( \mu_1^2\mu_2^2 + \mu_1^2\mu_3^2 + \mu_2^2\mu_3^2).
\end{equation}
And we have defined the forms \cite{Avramis:2007wb}
\begin{equation}
\mathcal{A}_1 = L^2( \mu_2^2 d\varphi_2 - \mu_3^2d\varphi_3 )~,
\qquad \mathcal{B}_1 = L^2\left[ -\mu_1^2 d\varphi_1 +
\ds\frac{\mu_2^2\mu_3^2}{\mu_2^2+\mu_3^2}(d\varphi_2+d\varphi_3)\right]
~,
\end{equation}
\begin{equation}
\mathcal{A}_2 = \mathcal{C}_1\wedge (d\varphi_1
+d\varphi_2+d\varphi_3)~, \qquad \mathcal{B}_2 = L^4 \left(
\mu_1^2\mu_2^2 d\varphi_1 \wedge d\varphi_2 + \mu_1^2\mu_3^2
d\varphi_3 \wedge d\varphi_1+ \mu_2^2\mu_3^2 d\varphi_2 \wedge
d\varphi_3\right) ~,
\end{equation}
\begin{equation}
\mathcal{A}_{4} = L^4  \ds\frac{r^4}{4} dt\wedge dx_1 \wedge dx_2
\wedge dy + \mathcal{C}_1 \wedge d\varphi_1 \wedge d\varphi_2
\wedge d\varphi_3\ ,
\end{equation}
where
\begin{equation}
d\mathcal{C}_{1} = L^4 \sin^3\vartheta \cos\vartheta
\sin\xi\cos\xi ~ d\vartheta \wedge d\xi .
\end{equation}
To facilitate the null Melvin twist we will need the explicit form
for the B-field
\begin{multline}
B_{(2)} = \gamma\mathcal{G}L^2 \left[ \mu_1^2\mu_2^2
d\varphi_1\wedge d\varphi_2 + \mu_1^2\mu_3^2 d\varphi_3\wedge
d\varphi_1 + \mu_2^2\mu_3^2 d\varphi_2\wedge d\varphi_3 \right]
\\- \sigma L^4 (w_1~ d\theta + w_2~d\xi)\wedge (d\varphi_1+
d\varphi_2 + d\varphi_3),
\end{multline}
where $w_i$ are defined as
\begin{equation}
\mathcal{C}_1 = L^4 ( w_1(\vartheta,\xi)~ d\vartheta +
w_2(\vartheta,\xi)~d\xi )~.
\end{equation}
Now make a coordinate change to the Hopf fiber coordinates
\begin{equation}
\vartheta = \mu~, \qquad \xi = \ds\frac{\alpha_1}{2}~, \qquad
\varphi_1 = \psi~, \qquad \varphi_2 = \psi +
\ds\frac{\alpha_3+\alpha_2}{2}~, \qquad \varphi_3 = \psi +
\ds\frac{\alpha_3-\alpha_2}{2}~, \label{Hopfcordchange}
\end{equation}
where $\alpha_i$ are the angles in the three $SU(2)$ left
invariant one forms (\ref{sigmas}). The metric becomes
\begin{multline}
ds^2_{\beta} = \mathcal{H}^{1/2}L^2 \left[ -r^2
\left(1-\ds\frac{r_{+}^4}{r^4}\right)dt^2 + r^2 (dx_1^2+dx_2^2
+dy^2) +
\left(1-\ds\frac{r_{+}^4}{r^4}\right)^{-1}\ds\frac{dr^2}{r^2}
\right] \\\\+ f_0' d\mu^2 + f_1' d\alpha_1^2 + f_2' d\alpha_2^2 +
f_3' d\alpha_3^2 + f_4'd\psi^2 + 2f_5' d\alpha_2d\alpha_3 + 2f_6'
d\psi d\alpha_3 + 2f_7' d\psi d\alpha_2.
\end{multline}
The B-field is
\begin{multline}
B_{(2)} = b_0' d\psi\wedge d\mu + b_1' d\psi\wedge d\alpha_1 +
b_2' d\psi\wedge d\alpha_2 + b_3' d\psi\wedge d\alpha_3 + b_4'
d\mu \wedge d\alpha_3 + b_5' d\alpha_1\wedge d\alpha_3 + b_6'
d\alpha_2 \wedge d\alpha_3.
\end{multline}
The dilaton is
\begin{equation}
\Phi = \ds\frac{1}{2} \ln (\mathcal{G}\mathcal{H}^2)\ ,
\end{equation}
where we have defined
\begin{eqnarray}
f_0' &=& L^2 \mathcal{H}^{1/2}~, \qquad f_1' = L^2
\mathcal{H}^{1/2} ~\ds\frac{\sin^2\mu}{4}~, \notag\\\notag\\
f_2'&=& L^2 \mathcal{H}^{1/2} \mathcal{G}~
\ds\frac{\sin^2\mu}{4}~, \qquad f_3' = L^2 \mathcal{H}^{1/2}
\mathcal{G}~ \left(\ds\frac{\sin^2\mu}{4} + |\beta|^2
L^4\mu_1^2\mu_2^2\mu_3^2 \right)~, \notag \\\\ \qquad f_4' &=& L^2
\mathcal{H}^{1/2} \mathcal{G}~ \left(1 + 9 |\beta|^2
L^4\mu_1^2\mu_2^2\mu_3^2 \right)~, \qquad f_5' = L^2
\mathcal{H}^{1/2} \mathcal{G}~
\ds\frac{\sin^2\mu\cos\alpha_1}{4}~, \notag\\\notag\\ f_6' &=& L^2
\mathcal{H}^{1/2} \mathcal{G}~ \left(\ds\frac{\sin^2\mu}{2} + 3
|\beta|^2 L^4\mu_1^2\mu_2^2\mu_3^2   \right)~, \qquad f_7' = L^2
\mathcal{H}^{1/2} \mathcal{G}~
\ds\frac{\sin^2\mu\cos\alpha_1}{2}~,\notag
\end{eqnarray}
and
\begin{eqnarray}
b_0' &=& 3 \sigma L^4 w_1~, \qquad b_1' = \ds\frac{3}{2} \sigma
L^4 w_2~, \qquad b_2' = \gamma \mathcal{G}L^4 \left(
\ds\frac{1}{2}\mu_1^2\mu_2^2 + \ds\frac{1}{2} \mu_1^2\mu_3^2 -
\mu_2^2\mu_3^2
\right)~,\notag\\\\
b_3' &=& \gamma \mathcal{G}L^4 ~ \ds\frac{1}{2} \mu_1^2 (\mu_2^2 -
\mu_3^2)~, \qquad b_4' = - \sigma L^4 w_1~, \qquad b_5' = -
\ds\frac{1}{2} \sigma L^4 w_2~,\qquad b_6' = \gamma \mathcal{G}
L^4~ \ds\frac{1}{2} \mu_2^2\mu_3^2~.\notag
\end{eqnarray}
%

\renewcommand{\theequation}{D.\arabic{equation}}
\setcounter{equation}{0}  
\section*{Appendix D. Hopf fiber of $S^5$}
\addcontentsline{toc}{section}{Appendix D. Hopf fiber of $S^5$}

The standard metric on the unit radius $S^5$ is
\begin{equation}
ds^2_{S^5} = \ds\sum_{i=1}^{3} \left(d\mu_i^2 +
\mu_i^2d\varphi_i^2\right) = d\vartheta^2 + \cos^2\vartheta
d\varphi_1^2 + \sin^2\vartheta (d\xi^2 + \cos^2\xi d\varphi_2^2 +
\sin^2\xi d\varphi_3^2)~,
\end{equation}
where we have defined
\begin{equation}
\mu_1 = \cos\vartheta~, \qquad \mu_2 = \sin\vartheta \cos\xi~,
\qquad \mu_3 = \sin\vartheta\sin\xi.
\end{equation}
After the coordinate change (\ref{Hopfcordchange}) the metric on
$S^5$ is written as a Hopf fiber over $\mathbb{CP}^2$
\begin{equation}
ds^2_{S^5} = ds^2_{\mathbb{CP}^2} + (d\psi+\mathcal{A})^2 = d\mu^2
+\ds\frac{\sin^2\mu}{4} \left(\sigma_1^2 + \sigma_2^2 + \cos^2\mu~
\sigma_3^2\right) + \left(d\psi +
\ds\frac{\sin^2\mu}{2}~\sigma_3\right)^2 ~,
\end{equation}
where $\sigma_i$ are defined in (\ref{sigmas}). Note that the
K\"{a}hler form on $\mathbb{CP}^2$ is
\begin{equation}
J = \ds\frac{1}{2} ~d\mathcal{A} = \ds\frac{1}{2}~\sin\mu\cos\mu ~
d\mu \wedge \sigma_3 + \ds\frac{1}{4} ~\sigma_1\wedge \sigma_2~,
\qquad \text{with} \qquad \mathcal{A}
=\ds\frac{\sin^2\mu}{2}~\sigma_3 ~.
\end{equation}
%

\renewcommand{\theequation}{E.\arabic{equation}}
\setcounter{equation}{0}  
\section*{Appendix E. T-duality rules}
\addcontentsline{toc}{section}{Appendix
E. T-duality rules}

Here we summarize the T-duality transformation rules for type II
theories with non-zero RR-flux \cite{Buscher:1987sk,Myers:1999ps}.
We assume that the T-duality is performed along the $y$-direction.

The NS fields transform under this according to
\begin{eqnarray}
\widetilde{g}_{yy} &=&  \ds\frac{1}{g_{yy}}~, \qquad
\widetilde{g}_{ay} =
\ds\frac{B_{ay}}{g_{yy}}~, \qquad \widetilde{g}_{ab} = g_{ab} - \ds\frac{g_{ay}g_{yb}+B_{ay}B_{yb}}{g_{yy}} ~,\notag\\\notag\\
\widetilde{B}_{ay} &=&  \ds\frac{g_{ay}}{g_{yy}}~, \qquad \widetilde{B}_{ab} = B_{ab} - \ds\frac{g_{ay}B_{yb}+B_{ay}g_{yb}}{g_{yy}} ~,\\\notag\\
\widetilde{\Phi} &=& \Phi - \ds\frac{1}{2}\log g_{yy} ~.\notag
\end{eqnarray}
The RR fields transform as
\begin{eqnarray}
&&\tilde{C}_{\mu\ldots\nu\alpha y}^{(n)}=C_{\mu\ldots\nu\alpha}^{(n-1)}-(n-1)\frac{C_{[\mu...\nu|y}^{(n-1)}g_{|\alpha]y}}{g_{yy}}\ ,\nonumber\\
&&
\tilde{C}_{\mu\ldots\nu\alpha\beta}^{(n)}=C_{\mu\ldots\nu\alpha\beta
y}^{(n+1)}+nC_{[\mu\ldots\nu\alpha}^{(n-1)}B_{\beta]y}+n(n-1)\frac{C_{[\mu\ldots\nu|y}^{(n-1)}B_{|\alpha|y}g_{|\beta]y}}{g_{yy}}\
.
\end{eqnarray}

It is also useful to write down the T-duality rules for the
RR-fluxes
\begin{eqnarray}
&& \tilde{F}_{\mu_1\ldots \mu_{n-1}y}^{(n)}=F_{\mu_1\ldots \mu_{n-1}}^{(n-1)}+(n-1)(-1)^n\frac{g_{y[\mu_1}F_{\mu_2\ldots \mu_{n-1}]y}^{(n-1)}}{g_{yy}}\ ,\nonumber\\
&& \tilde{F}_{\mu_1\ldots \mu_{n}}^{(n)}=F_{\mu_1\ldots
\mu_{n}y}^{(n+1)}-n(-1)^nB_{y[\mu_1}F_{\mu_2\ldots
\mu_n]}^{(n-1)}-n(n-1)\frac{B_{y[\mu_1}g_{\mu_2|y|}F_{\mu_3\ldots
\mu_n]y}^{(n-1)}}{g_{yy}}\ .
\end{eqnarray}



\end{document}